\documentclass[12pt]{article}

\usepackage{mathptmx}       
\usepackage{helvet}         
\usepackage{courier}        
\usepackage[american]{babel}
\usepackage{latexsym,amssymb}
\usepackage{txfonts}

\makeatletter
\renewcommand{\@evenhead}{\raisebox{0pt}[\headheight][0pt]{\vbox{\hbox
to \textwidth{\thepage\hfil\strut\textit{\leftmark}}\hrule}}}
\renewcommand{\@oddhead}{\raisebox{0pt}[\headheight][0pt]{\vbox{\hbox
to \textwidth{\textit{\rightmark}\hfil\strut\thepage}\hrule}}}
\makeatother

\def\a{\alpha}
\def\b{\beta}

\def\om{\omega}
\let\m=\mu
\let\n=\nu
\def\na{\nabla} 
\def\II{{\mathbb{I}}} 
\def\RR{{\mathbb{R}}} 
\def\CC{{\mathbb{C}}}

\def\R{\mathcal{R}} 
\def\F{\mathcal{F}} 
\def\tr{\mathrm{tr\,}} 
\def\Tr{\mathrm{Tr\,}} 
\def\Det{\mathrm{Det\,}} 
\def\vol{\mathrm{vol\,}} 
 
\def\diag{\mathrm{diag\,}} 
\def\Spin{\mathrm{Spin\,}} 
\def\End{\mathrm{End\,}} 

\def\be{\begin{equation}}
\def\ee{\end{equation}}
\def\bea{\begin{eqnarray}}
\def\eea{\end{eqnarray}}
\def\bes{\begin{displaymath}}
\def\ees{\end{displaymath}}
\def\bmp{\begin{minipage}}
\def\emp{\end{minipage}}
\newtheorem{theorem}{Theorem}

\begin{document}

\begin{titlepage}
\thispagestyle{empty}
\null
\hspace*{50truemm}{\hrulefill}\par\vskip-4truemm\par
\hspace*{50truemm}{\hrulefill}\par\vskip5mm\par
\hspace*{50truemm}{{\large\sc 
New Mexico Tech {\rm (December 17, 2008)
}}}\vskip4mm\par
\hspace*{50truemm}{\hrulefill}\par\vskip-4truemm\par
\hspace*{50truemm}{\hrulefill}
\par
\bigskip
\bigskip
\vspace{3cm}
\centerline{\huge\bf Mathematical Tools}
\bigskip
\centerline{\huge\bf for Calculation of the
Effective Action}
\bigskip
\centerline{\huge\bf in Quantum Gravity}
\bigskip
\bigskip
\centerline{\Large\bf Ivan G. Avramidi}
\bigskip
\centerline{\it Department of Mathematics}
\centerline{\it New Mexico Institute of Mining and Technology}
\centerline{\it Socorro, NM 87801, USA}
\centerline{\it E-mail: iavramid@nmt.edu}
\bigskip
\medskip
\vfill

{\narrower
\par

We review the status of covariant methods in quantum field theory and quantum
gravity, in particular, some recent progress in the calculation of the
effective action via the heat kernel method. We study the heat kernel
associated with an elliptic second-order partial differential operator of
Laplace type acting on smooth sections of a vector bundle over a Riemannian
manifold without boundary. We develop a manifestly covariant method for
computation of the heat kernel asymptotic expansion as well as new algebraic
methods for calculation of the heat kernel for covariantly constant background,
in particular, on homogeneous bundles over symmetric spaces, which enables one
to compute the low-energy non-perturbative effective action.

\par}
\vfill

\end{titlepage}

\section{Introduction}
\setcounter{equation}0

One of the most important problems of modern fundamental physics is the problem
of reconciling classical general relativity, the theory of macroscopic
gravitational phenomena, with quantum theory, so-called Quantum Gravity
problem. This is a really difficult task since one has to answer the very basic
questions concerning the local and the global structure of the spacetime itself
as well as deep questions about the nature of quantum mechanics. 

Although, over the last several decades many competing approaches (Euclidean
path integrals, string theory, loop gravity, non-commutative geometry,
asymptotic safety, various lattice approaches and others) has been put forward
and despite some real progress in some of these approaches in the last two
decades, we still do not have a complete consistent theory of quantum
gravitational phenomena. It looks like we are missing an important piece
of the puzzle which prevents us to find the solution.

In this situation it seems to be wise to go back and to recall some pioneering
works in quantum gravity. This review will concentrate on so-called
{\it covariant
methods} in quantum gravity. Some other approaches are reviewed by other
lecturers of this school. The basis of the covariant methods in quantum gravity
is the background field method. This method was developed mainly by De~Witt in
his classical papers \cite{dewitt67a, dewitt67b} and reviews \cite{dewitt75,
dewitt79} (for the latest update see the book \cite{dewitt03}). It is a
generalization of the method of generating functionals in quantum field theory
developed and successfully used by Schwinger \cite{schwinger51,schwinger54}.
For a detailed review see, for example,
\cite{buchbinder92,birrel82,fulling89}). 

The basic object in the background field method is the {\it effective action}.
The
effective action is a functional of the background fields that encodes, in
principle, all the information of quantum field theory.
It determines the full
one-point propagator and the full vertex functions and, hence, the whole
$S$-matrix. Moreover, the variation of the effective action gives the effective
equations for the background fields, which makes it possible to study the 
back-reaction of quantum processes on the classical background. In particular,
the low energy effective action (called the effective potential) is the most
appropriate tool for investigating the structure of the physical vacuum in
quantum field theory.

The only practical method for the calculation of the effective action is the
semi-classical
perturbative expansion of the path integral in the number of loops. All fields
are split in background classical parts and quantum perturbations propagating
on this background and the classical action is expanded in quantum fields. Then
the quadratic part determines the propagators of the quantum fields and the
higher-order terms reproduce the vertex functions of the perturbation theory.

In the perturbation theory the effective action is expressed in terms of the
propagators and the vertex functions. One of the most powerful methods to study
the propagators is the {\it proper time method}
(also called the {\it heat kernel method},
in particular, by mathematicians), which was originally proposed by Fock
\cite{fock37} and later generalized by Schwinger 
\cite{schwinger51,schwinger54}
who also
applied it to the calculation of the one-loop  effective action in quantum
electrodynamics. It was De Witt \cite{dewitt67a, dewitt75, dewitt79} who
perfected the proper time method; he   reformulated it in the geometrical
language and applied it to the case of gravitational field.

At   one-loop level, the contribution of the  gravitational loop is of the same
order as the contributions of matter fields . At low energies (lower than the 
Planckian energy, $\hbar c^5/G$) the contribution of higher gravitational loops
should be highly suppressed. Therefore, a semi-classical concept applies when
the quantum matter fields together with the linearized perturbations of the
gravitational field interact with the background  gravitational field (and,
probably, with the background  matter fields). This is what is usually called
the {\it one-loop quantum gravity}. 
The main difficulty of  quantum gravity is
the
fact that there is no consistent way to eliminate the ultraviolet divergences
arising in perturbation theory, even at one loop level.

The present review is devoted to the development of the covariant methods for
calculation of the effective action in quantum field theory and quantum
gravity. In Sect.2 we review the formal structure of quantum gauge field theory
and quantum gravity and the construction of the effective action following
\cite{avramidi00,avramidi02}. In Sect. 3 we describe the heat kernel method and
develop the asymptotic expansion of the heat kernel following
\cite{avramidi90b,avramidi91,avramidi99a,avramidi00,avramidi02}. In Sect. 4 we
describe the local structure of the Green function following \cite{avramidi98}.
In Sect. 5 we develop a method for the calculation of the heat kernel
coefficients and describe their general structure following
\cite{avramidi91,avramidi99a,avramidi00,avramidi02}. In Sect. 6 we compute the
heat trace in the high-energy approximation following
\cite{avramidi90a,avramidi91,avramidi00}. In Sect. 7 we describe our results
for the calculation of the low-energy heat trace following our recent work
\cite{avramidi93,avramidi94a,avramidi95c,avramidi96,avramidi08a,avramidi08b}.
In Sect. 8 we apply the obtained results to compute the low-energy one-loop
effective action in quantum gravity.

\section{Effective Action in Quantum Field Theory and
\\
Quantum Gravity}
\setcounter{equation}0


In this section we briefly describe the standard formal construction of the
generating  functional and the effective action in gauge theories. The basic
object of any physical theory is the spacetime $M$, which we will assume  to be
a  $n$-dimensional manifold with the topological structure of a cylinder  
\be
M =I\times\Sigma\,,
\ee
where $I$ is an open interval of the  real line (or the whole  real line) and
$\Sigma$ is some $(n-1)$-dimensional manifold. The spacetime manifold is here
assumed to be globally hyperbolic and equipped with a (pseudo)-Riemannian
metric $g$ of signature $(- + \cdots +)$; thus, a foliation of spacetime exists
into spacelike sections identical to $\Sigma$. Usually one also assumes the
existence of a spin structure on $M$. A point $x=(x^\mu)$ in the spacetime is
described locally by the time  $x^0$ and the space coordinates $(x^1, \dots,
x^{n-1})$. We label the spacetime  coordinates by Greek indices, which run from
$0$ to  $(n-1)$, and sum over  repeated indices. 

Let us consider a vector bundle ${\cal V}$ 
over the spacetime $M$ each fiber of which
is isomorphic to a vector space, $V$,
on which the spin group ${\rm Spin}(1,n-1)$,
i.e. the covering  group of Lorentz group,  acts. The vector bundle
${\cal V}$ can
also have an additional structure on which a  gauge group acts. The sections of
the vector bundle 
${\cal V}$ are called fields. The  tensor fields describe the
particles with integer spin (bosons) while the spin-tensor fields  describe
particles with half-integer spin (fermions).  Although the whole scheme can be
developed for superfields (a combination  of boson and fermion fields), we
restrict ourselves in  the present lecture to
boson fields (which, without loss of generality can be considered real).  
A field $\varphi$ 
is represented locally by a set of real-valued functions 
$
\varphi=(\varphi^A(x))\,,
$
where
$A=1,\dots, \dim V$. Capital Latin indices will be used to label the local
components of the fields. To construct invariant functionals we need
to introduce an invariant fiber inner product
and an $L^2$ inner product
\be
(\psi,\varphi)=\int\limits_M d\vol(x)\,\psi^A(x)
E_{AB}(x)\varphi^B(x)\,.
\ee
where $d\vol(x)=dx\,g^{1/2}$, $g=|\det g_{\mu\nu}|$, is the natural
Riemannian volume element defined by some background  metric $g$, and
$E^{AB}$ is a non-degenerate symmetric matrix (a fiber
metric).
As usual, we assume that a summation over repeated
indices is performed.
This metric (and its inverse $E^{-1}{}^{AB}$) can be used to naturally
identify the bundle 
${\cal V}$ with its dual ${\cal V}^*$ (that is to raise and lower the
field indices). The sections of the dual bundle are called currents and are
represented locally by a set of functions, e.g.
\be
J_A=E_{AB}\varphi^B\,.
\ee
We will also use the condensed De Witt notation,
where the discrete index $A$ and the spacetime point $x$ are combined in one 
lower case Latin index 
$
i \equiv (A, x)\,.
$
Then the components of a field $\varphi$ are $(\varphi^{i}) \equiv
(\varphi^{A}(x))$. There is a natural pairing between the bundles 
${\cal V}$ and
${\cal V}^*$ defined by
\be
\left<J,\varphi\right>\equiv
J_{i}\varphi^{i}\equiv
\int\limits_{M}d\vol(x)\, J_{A}(x)\varphi^{A}(x)\,.
\ee
It is assumed that a summation over repeated
lower case Latin indices, i.e. a combined  summation-integration, is 
performed. 

The set of all sections of the vector bundle 
${\cal V}$ is called  the configuration
space, which one assumes to be an
infinite-dimensional
manifold ${\cal M}$.  The fields $\varphi^i$ are the coordinates on this
manifold, the variational derivative $\delta/\delta\varphi$ is a tangent
vector, a small disturbance $\delta\varphi$ is a one-form and so on. If
$S(\varphi)$ is a scalar field on the configuration space, then its variational
derivative $\delta S/\delta\varphi$ is a one-form on ${\cal M}$  defined by
\be
\frac{d}{d\varepsilon}S(\varphi+\varepsilon h)\Big|_{\varepsilon=0}
=\left<\frac{\delta S}{\delta\varphi},h\right>
=\frac{\delta S}{\delta\varphi^i}h^i\,.
\ee
By using the functional 
differentiation one can define formally the concept of tangent space,  the
tangent vectors, Lie derivative, one-forms, metric, connection,
geodesics and so on (for more details, see \cite{dewitt03}).

\subsection{Non-Gauge Field Theories}

The dynamics of quantum field theory is determined by an action functional
$S(\varphi)$, which is a differentiable real-valued scalar field on the
configuration space.  The dynamical field configurations are defined as the 
field configurations  satisfying the stationary action principle, i.e. they
must  satisfy the dynamical equations of motion 
\be
\frac{\delta S}{\delta\varphi}=0
\ee
with given boundary (and initial) conditions. The set of all dynamical  field
configurations, i.e. those that satisfy the dynamical equations  of motion,
${\cal M}_{0}$, is a subspace of the configuration space  called the dynamical
subspace (or the mass-shell in the high-energy physics jargon). 

Quantum field theory is  basically a theory of small disturbances on the
dynamical subspace.  Most of the problems of standard quantum field theory deal
with scattering processes, which are described by the transition amplitudes
between some well defined initial  and final states in the remote past and the
remote future.  The collection of all these amplitudes is called  the
scattering matrix, or shortly $S$-matrix. 


Let us single out in the space-time two causally connected in-- and out--
regions, that lie in the past and in the future respectively relative to the
region $\Omega$, which is of interest from the dynamical standpoint.  Let
$\left.|{\rm in}\right>$ and $\left.|{\rm out}\right>$ be some initial and
final states of the quantum field system in these regions. Let us consider the
transition amplitude $\left<{\rm out|in}\right>$ and ask the question: how does
this amplitude change under a variation of the interaction with a compact
support in the region $\Omega$. The answer to this question gives the 
{\it Schwinger
variational principle} which states that
\be
\delta\left<{\rm out}|{\rm in}\right> 
= {i\over\hbar}\left<{\rm out}|\right.\delta S\left.|{\rm in}\right>,
\ee
where $\delta S$ is the corresponding change of the action. This principle
gives a very powerful tool to study the transition amplitudes. The Schwinger
variational principle can be called the 
{\it quantization postulate}, because all the
information about quantum fields can be derived from it.

Let us change the external conditions by adding a linear interaction 
with  some external classical sources $J$ in the dynamical 
region $\Omega$, i.e. 
\be
\delta S=\left<J,\varphi\right>\,.
\ee
The amplitude $\left<{\rm out}|{\rm in}\right>$ 
becomes a functional of the sources that we denote by $Z (J)$. 
The primary objects of interest in quantum field theory are
the chronological mean values 
\bea
\Psi_n^{i_{n}\dots i_{1}}
&\equiv&
\frac{\left<{\rm out}|T(\varphi^{i_{n}}
\cdots\varphi^{i_{1}})|{\rm in}\right>}
{\left<{\rm out}|{\rm in}\right>}
\eea
where $T$ denotes the operator of chronological ordering that orders the
(non-commuting) operators in order of their time variables from right to left.
Of course, in the presence of the sources they become functionals of $J$.
By using the Schwinger variational principle one can obtain
the chronological mean values
in terms of the functional derivatives 
of the functional $Z (J)$, that is,
\bea
Z(J+\eta)&=&Z(J)
\Biggl\{1
+\left(\frac{i}{\hbar}\right)
\left<\eta,\Psi_1(J)\right>
+\frac{1}{2}\left(\frac{i}{\hbar}\right)^2
\left<\eta,\Psi_2(J)\eta\right>
\nonumber\\
&&+
\sum_{n=3}^\infty \frac{1}{n!}\left(\frac{i}{\hbar}\right)^n
\Psi_n^{i_{n}\dots{i_1}}(J)\eta_{i_1}\cdots \eta_{i_n}
\Biggr\}\,.
\eea
In other words, the functional $Z (J)$ is the generating functional for the
chronological amplitudes $\Psi_n$. 

Let us now define another functional $W(J)$ by
\be
Z=\exp\left({i\over\hbar}W\right)\,.
\ee
Its functional derivatives define so called full 
{\it connected Green functions}, 
${\cal G}_n^{i_{i}\ldots i_{n}}$, (or the 
{\it correlation functions}) by
\bea
W(J+\eta)=W(J)+\left<\eta,{\cal G}_1(J)\right>
+\frac{1}{2}\left<\eta, {\cal G}_2(J)\eta\right>
+\sum_{n=3}^\infty \frac{1}{n!}
{\cal G}_n^{i_{i}\ldots i_{n}}(J)\eta_{i_1}\cdots \eta_{i_n}\,.
\eea
The functional $\phi={\cal G}_1$ is called the background (or the mean) field,
the operator ${\cal G}={\cal G}_2$ is
called  the full propagator.
Then, it is easy to see that all chronological mean amplitudes can be
expresssed in terms of connected Green functions.
In particular, we have
\bea
\Psi_1&=&\phi,
\\
\Psi^{jk}_2&=&\phi^j\phi^k+\frac{\hbar}{i}{\cal G}^{jk}\,.
\eea
Thus, whilst $Z(J)$ is the 
generating functional for
chronological amplitudes, the  functional $W(J)$ is the generating functional
for the connected  Green functions. The Green functions satisfy the boundary 
conditions which are determined by the states  $\left.|{\rm in}\right>$ and
$\left|{\rm out}\right>$.

The mean field itself is a functional of the sources, $\phi=\phi(J)$.
It is easy to see that 
the functional 
derivative of the mean field is equal to the full propagator,
that is,
\be
\frac{d}{d\varepsilon}\phi(J+\varepsilon\eta)\Big|_{\varepsilon=0}
={\cal G}\eta\,.
\ee
In the non-gauge theories the full propagator ${\cal G}$, 
which plays the role of the (infinite-dimensional) Jacobian,  
is non-degenerate. Therefore, one can change variables and 
consider $\phi$ as independent variable and $J=J(\phi)$ (as well as all other
functionals) as a functional of $\phi$.

There are many different ways to show that there is a functional
$\Gamma(\phi)$ such that  
\be
\left<\frac{\delta S(\varphi)}{\delta\varphi}\right>
=\frac{\delta\Gamma(\phi)}{\delta\phi}\,.
\ee
This functional is defined by
\be
\left<{\rm out}|{\rm in}\right>
=\exp\left\{{i\over \hbar}\left[\Gamma
+\left<J,\phi\right>\right]\right\},
\ee
or by the functional Legendre transform
\be
\Gamma(\phi)=W(J(\phi))-
\left<J(\phi),\phi\right>\,.
\ee

This is the most important object in quantum field theory.
It contains all the information
about quantized fields. 
The functional expansion of this functional reads
\be
\Gamma(\phi+h)
=\Gamma(\phi)
-\left<J(\phi),h\right>
-\frac{1}{2}\left<h,{\cal G}(J(\phi))h\right>
+\sum_{n=3}^\infty \frac{1}{n!}
\Gamma_{,i_1\dots i_n}(\phi)h^{i_n}\cdots h^{i_1}\,.
\ee
Therefore, the first variation of 
$\Gamma$ gives the effective equations for the background fields
\be
{\delta \Gamma\over\delta\phi}=-J\,.
\ee
These equations replace the classical equations of motion 
and describe the effective dynamics of the background field 
with regard to all quantum corrections.
That is why $\Gamma$ is called the {\it effective action}. 

Furthermore, 
the second derivative of $\Gamma(\phi)$ determines the full propagator 
\bes
{\cal G}=\left(-{\delta^2\Gamma\over\delta\phi^2}\right)^{-1}\,.
\ees
The higher derivatives, $\Gamma_{,i_1\cdots i_k}$, determine the so-called full
{\it vertex functions}
(also called {\it strongly connected}, or {\it one-particle irreducible},
functions). In other words, $\Gamma(\phi)$ is the generating functional  for
the full vertex functions. The full vertex functions  together with  the full
propagator determine the full connected Green functions and, therefore,  all 
chronological amplitudes and, hence, the $S$-matrix. Thus, the entire quantum
field theory is summed up in the functional structure  of the effective
action. 

One can obtain a very useful formal representation for the effective action in
terms of functional integrals (called also 
{\it path integrals}, or  Feynman
integrals). A functional integral is an integral over the
(infinite-dimensional) configuration space ${\cal M}$.  Although a rigorous
mathematical definition of functional integrals is absent, they can be
used in perturbation theory of quantum field theory as an  effective tool,
especially in gauge theories, for manipulating the  whole series of
perturbation theory. The point is that in perturbation  theory one encounters
only functional integrals of Gaussian type, which can be well defined
effectively in terms of classical propagators and vertex functions. The
Gaussian integrals do not depend much on the dimension and, therefore, (after a
proper normalization) all formulas from the finite-dimensional case, like
Fourier transform, integration by parts, delta-function, change  of variables
etc., are valid in the infinite-dimensional case as well. One has to note that
functional integrals are formally divergent --- if one tries to evaluate
the integrals, one encounters meaningless divergent expressions. This
difficulty can be overcome in the framework of the renormalization theory (in
so-called renormalizable field theories). In non-renormalizable theories (like
quantum general relativity) this issue becomes the main difficulty of the
theory.

Integrating the Schwinger variational principle one can obtain the
following functional integral:
\be
\left<{\rm out}|{\rm in}\right>=
\int\limits_{\cal M}{\cal D}\varphi\;
\exp\left\{{i\over\hbar}\left[S(\varphi)
+\left<J,\varphi\right>\right]\right\}.
\ee
Here ${\cal D}\varphi$ represents the functional measure; however,
it should not be taken too seriously---it will just provide a formal
device for manipulations of Gaussian integrals. 
Correspondigly, for the effective action one obtains 
the functional equation
\bea
\exp\left\{{\frac{i}{\hbar}\Gamma(\phi)}\right\}
&&=
\int\limits_{\cal M}{\cal D}\varphi\;\exp\left\{{\frac{i}{\hbar}
\left[S(\varphi)
-\left<{\delta\Gamma(\phi)\over\delta\phi},
(\varphi-\phi)\right>\right]}\right\}\,.
\eea

The only way to get numbers from this formal 
expression is to take advantage of the semi-classical 
approximation within a formal (asymptotic) 
expansion in powers of Planck
constant $\hbar$: 
\be
\Gamma\sim
S+\sum\limits_{k=1}^\infty \hbar^{k}\Gamma_{(k)}.
\ee
Next, we 
substitute this expansion in the functional equation for the 
effective action, shift the integration variable in the 
functional integral 
\be
\varphi=\phi+\sqrt\hbar\,h\,,
\ee
and expand the action $S(\varphi)$ in functional Taylor 
series in quantum fields $h$
\bea
S(\phi+\sqrt \hbar\,h)
&=&S(\phi)
+\hbar^{1/2}\left<\frac{\delta S(\phi)}{\delta\phi},h\right>
-\hbar\frac{1}{2}\left<h,L(\phi)h\right>
\nonumber\\
&&
+\sum_{n=3}^\infty \frac{1}{n!}\hbar^{n/2} 
S_{,i_1\dots i_n}(\phi)h^{i_n}\cdots
h^{i_1}\,,
\eea
where $L$ is a (usually, partial differential)
operator defined by the second
variation of the action
\be
L=-{\delta^2 S\over\delta\varphi^2}\,.
\ee 
Notice that the operator $L$ maps sections of the vector bundle 
${\cal V}$ to sections
of the dual bundle ${\cal V}^*$, that is,
\be
L: C^{\infty}({\cal V})\to C^\infty({\cal V}^*)\,.
\ee
In order to have a well defined operator which is 
self-adjoint with respect
to the $L^2$ inner product on the bundle 
${\cal V}$ we define another operator 
\be
\hat L: C^\infty({\cal V})\to C^{\infty}({\cal V})\,,
\ee
such that
\be
(\varphi,\hat L h)
=\int\limits_M d\vol\;
\varphi^A E_{AB} \hat L^B{}_C h^C
=\int\limits_M d\vol\;
\varphi^A L_{AC} h^C
=\left<\varphi,Lh\right>  \,,
\ee
that is,
\be
E_{AB} \hat L^B{}_C
=L_{AC} h^C\,.
\ee

Now, by 
expanding both sides of 
the functional equation for the effective action 
in powers on $\hbar$   and
equating the 
coefficients of equal powers of $\hbar$, we get the recurrence
relations that uniquely define all coefficients $\Gamma_{(k)}$.
The measure formally transforms as
$
{\cal D}\varphi={\cal D}h\,.
$ 
All functional integrals appearing in this expansion are 
Gaussian and can be calculated in terms of the functional
determinant, $\Det \hat L$, of the operator $\hat L$ and  the bare propagator
$G=L^{-1}$,
i.e. the Green function of the operator  $L$ with Feynman boundary
conditions.
More precisely, with the proper normalization of the measure one can define
\bea
\int\limits_{\cal M}{\cal D}h\;\exp\left(-\frac{i}{2}
\left(h,\hat L h\right)\right)
&=&(\Det \hat L)^{-1/2}\,,
\\
\int\limits_{\cal M}{\cal D}h\;
\exp\left(-\frac{i}{2}\left(h,\hat L h\right)
\right)h^{i_{1}}\cdots h^{i_{2m+1}}
&=&0\,,
\\
\int\limits_{\cal M}{\cal D}h\;
\exp\left(-\frac{i}{2}\left(h,\hat L h\right)
\right)h^{i_{1}}\cdots h^{i_{2m}}
&=&\frac{(2m)!}{2^m m! i^m}
(\Det \hat L)^{-1/2}G^{(i_1i_2}\cdots G^{i_{2m-1}i_{2m})}
\,,
\nonumber\\
&&
\eea
where parenthesis denote the complete symmetrization over all indices
included.
Of course, the Green functions of the operators $L$ and $\hat L$ are
related by
\be
\hat G^A{}_B(x,y)=G^{AC}(x,y)E_{CB}(y)\,.
\ee

In particular, 
the one-loop effective action is determined by the functional 
determinant of the operator $L$ 
\be
\Gamma_{(1)}=-\frac{1}{2i}\log\Det\,\hat L\,,
\ee
and the two-loop effective action is given by
\be
\Gamma_{(2)}=
-\frac{1}{8}S_{,ijkl}G^{ij}G^{kl}
-\frac{1}{12}
S_{,ijk}G^{il}G^{jm}G^{kn}
S_{,lmn}\,.
\ee

Strictly speaking,  the Gaussian integrals are well defined for elliptic
partial differential operators  in terms of the functional determinants and
their Green functions.  Although the Gaussian integrals of quantum field theory
are determined  by hyperbolic partial differential operators with Feynman
boundary conditions  they can be well defined by means of the analytic 
continuation from the  Euclidean sector of the theory where the operators
become elliptic. This is done by so-called Wick rotation---one replaces the
real time coordinate by a purely imaginary  one $x^0\to i\tau$ and singles out
the imaginary factor also from the  action $S\to iS$ and the effective action
$\Gamma\to i\Gamma$.  Then the metric of the spacetime manifold becomes
positive  definite and the classical action in all `nice' field theories
becomes a positive-definite functional. Then the fast oscillating Gaussian
functional integrals become exponentially decreasing and can  be given a
rigorous mathematical meaning. 

\subsection{Gauge Field Theories}

Let us try to apply the formalism described above to a gauge field theory.
A characteristic feature of a gauge field theory is the fact that 
the dynamical equations 
\be
\frac{\delta S}{\delta\varphi}=0
\ee
are not independent 
--- there  are certain identities, called N\"other identities,  between them.
This means that there are some nowhere vanishing vector fields
\be
\RR_\alpha=R^i{}_\alpha\frac{\delta}{\delta\varphi^i}
\ee
on the configuration space 
${\cal M}$ that annihilate the action,
\be
{\RR}_{\alpha}S=0\,,
\ee
and, hence,
define invariance flows on ${\cal M}$. The transformations of the fields 
\be
\delta_{\xi}\varphi^i=R^i{}_{\alpha}\xi^\alpha
\ee
are called the invariance
transformations and $\RR_{\alpha}$ are called the generators of invariance
transformations. The infinitesimal parameters of these transformations $\xi$
are sections of another vector bundle (usually the tangent bundle $TG$ of a
compact Lie group $G$) that are respresented locally by a set of functions
$
(\xi^\alpha)=(\xi^a(x))\,,
$
$a=1,\dots,\dim G$,  over spacetime with compact
support.  To distinguish between the components of the gauge fields and the
components of the gauge parameters we introduce  lower case Latin indices from
the beginning of the alphabet; the Greek indices from the beginning of the
alphabet are used as condensed labels 
$
\alpha =(a,x)
$
that include the spacetime point.  

We assume that the vector fields $\RR_{\alpha}$ are linearly
independent and complete, which means that they form a complete basis in the
tangent space of the invariant subspace of configuration space. The vector
fields $\RR_{\alpha}$ form the gauge algebra. We restrict ourselves to the
simplest case when the gauge algebra is the Lie algebra of an
infinite-dimensional gauge Lie group $\cal G$.
This is the case in Yang-Mills theory and gravity. Then the flow vectors
$\RR_{\alpha}$ decompose the configuration space into the invariants subspaces
of ${\cal M}$ (called the orbits) consisting of the points connected by the
gauge transformations. The space of orbits is then ${\cal M}/{\cal G}$.  The
linear independence of the vectors $\RR_{\alpha}$  at each point implies that
each orbit is a copy of the group manifold. One can show that the vector fields
$\RR_{\alpha}$ are tangent to the dynamical subspace ${\cal M}_0$, which means
that the orbits do not intersect ${\cal M}_0$ and the invariance flow maps the
dynamical subspace ${\cal M}_0$ into itself. Since all  field configurations
connected by a gauge transformation, i.e., the points on an orbit, are
physically  equivalent, the physical dynamical variables  are the classes of
gauge equivalent field configurations, i.e.,  the orbits. The physical
configuration space is, hence, the  space of orbits ${\cal M}/{\cal G}$. In
other words  the physical observables must be the invariants of the  gauge
group.

To quantize a gauge theory by means of the functional integral,  we consider
the in-- and out-- regions,  define some $\left.|{\rm in}\right>$ and
$\left.|{\rm out}\right>$  states in these regions and study the amplitude
$\left<{\rm out}|{\rm in}\right>$. Since all field configurations along  an
orbit are  physically equivalent we have to integrate  over the orbit space
${\cal M}/{\cal G}$.  To deal with such situations one has to choose a
representative  field in each orbit. This can be done by choosing special
coordinates  $(I^A(\varphi),\chi^{\alpha}(\varphi))$ on the configuration space
 ${\cal M}$, where $I^A$ label the orbits and $\chi^{\alpha}$ the points in the
orbit. Computing the Jacobian of the field transformation and introducing a
delta functional $\delta(\chi-\zeta)$ we can fix the coordinates on the orbits
and obtain the measure on the orbit space ${\cal M}/{\cal G}$  
\be
{\cal D}I={\cal D}\varphi\,\Det F(\varphi)\delta(\chi(\varphi)-\zeta),
\ee
where 
\be
F^\beta{}_\alpha=\RR_\alpha \chi^\beta
\ee
is a non-degenerate operator. 
Thus we obtain a functional integral for the transition amplitude
\bea
\left<{\rm out}|{\rm in}\right>&=&\int\limits_{\cal M}{\cal D}\varphi\;
\Det F(\varphi)\delta(\chi(\varphi)-\zeta)
\exp\left\{{i\over\hbar}S(\varphi)\right\}.
\eea

Now one can go further and integrate this equation over parameters $\zeta$ with
a Gaussian measure determined by a symmetric nondegenerate matrix
$\gamma=(\gamma_{\alpha\beta})$, which most naturally can be 
chosen as the
metric on the orbit (gauge group metric). As a result we get 
\bea
\left<{\rm out}|{\rm in}\right>
=\int\limits_{\cal M}{\cal D}\varphi\,
\left(\Det\gamma\right)^{1/2}\,\Det F(\varphi)
\exp\left\{{i\over\hbar}\left[S(\varphi)
+\frac{1}{2}\left<\chi(\varphi),\gamma\chi(\varphi)\right>\right]\right\}.
\eea
The functional equation for the effective action takes the form
\bea
&&\exp\left\{{i\over \hbar}\Gamma(\phi)\right\}
=
\int\limits_{\cal M} {\cal D}\varphi\,
\left(\Det\gamma(\phi)\right)^{1/2}\,
\Det F(\varphi)
\\
&&\qquad\qquad
\times \exp\Biggl\{{i\over \hbar}\left[S(\varphi) 
+{1\over 2}\left<\chi(\varphi),\gamma(\phi)\chi(\varphi)\right>
-\left<{\delta\Gamma(\phi)\over\delta\phi},
(\varphi-\phi)\right>\right]\Biggr\}\,.
\nonumber
\eea
The determinants of the operators $F$ and $\gamma$ are usually
represented as a result of the integration over some auxiliary
Grassmanian variables, so called ghost fields. 

This equation can be used to construct the 
semi-classical perturbation theory
in powers of the Planck constant (loop expansion), which gives the
effective action in terms of the bare propagators and the vertex functions.
In particular, one finds the one-loop effective action  
\bea
\Gamma_{(1)} &=& -{1\over 2i }\log\Det\;\hat L
+{1\over i}\log\Det F
+{1\over 2i}\log\Det\gamma\;,
\eea
where $\hat L$ is an operator defined by
\be
\frac{d^2}{d\varepsilon^2}\left\{
S(\varphi+\varepsilon h) 
+{1\over 2}\left<\chi(\varphi+\varepsilon h),
\gamma\chi(\varphi+\varepsilon h)\right>\right\}\Big|_{\varepsilon=0}
=-(h,\hat L h)\,.
\ee
In De Witt notation it reads
\be
\hat L^k{}_j=E^{-1}{}^{ki}L_{ij}\,,
\ee
where
\be
L_{ij}=
-{\delta^2 S\over\delta\varphi^i\delta\varphi^j} 
-{\delta\chi^\alpha\over\delta\varphi^i}\gamma_{\alpha\beta}
{\delta\chi^\beta\over\delta\varphi^j}\,.
\ee


\subsection{Quantum General Relativity}

Einstein's theory of general relativity is an example of a gauge theory with
the gauge group ${\cal G}$ being the group of all diffeomorphisms of the
spacetime manifold $M$ and the configuration space ${\cal M}$ being the space
of all pseudo-Riemannian metrics  on $M$. The physical configuration space
${\cal M}/{\cal G}$ of all orbits of the gauge group is then the space of all
geometries on the spacetime.

The  gravitational field can be parametrized by the 
metric tensor of the space-time
\be
\varphi^i  = g_{\mu\nu}(x)\;,
\qquad i\equiv (\mu\nu, x)\;.
\ee
An invariant fiber metric is defined by
\be
E^{\mu\nu\alpha\beta}=g^{\mu(\alpha}g^{\beta)\nu}
-\varkappa g^{\mu\nu}g^{\alpha\beta}\,,
\ee
where
$\varkappa\ne 1/n$ is a real parameter.
The inverse metric is then 
\be
E^{-1}_{\mu\nu\alpha\beta}=g_{\mu(\alpha}g_{\beta)\nu}
-\frac{\varkappa}{n\varkappa-1}g_{\mu\nu}g_{\alpha\beta}\,.
\ee

The parameters of gauge
transformations are the components of the vector of the
infinitesimal diffeomorphism, 
\be
\xi^\mu  = \xi^\mu(x)\;, \qquad \mu\equiv (\mu,x)\;.
\ee
An invariant metric in the gauge group can be chosen to be
just a background metric $g_{\mu\nu}$.

The local  generators of the gauge transformations in this
parametrization are defined by their action as follows
\be
R^i{}_{\alpha}\xi^\alpha
=2\nabla_{(\mu}\xi_{\nu)}\;, 
\qquad i\equiv(\mu\nu,x)\;,
\ee
\be
J_i R^i{}_{\alpha}
=-2\nabla_{\mu}J^{\mu}{}_{\alpha}\;, 
\qquad \alpha\equiv(\alpha,x)\;.
\ee

The Hilbert-Einstein action of general relativity
has the form  
\be
S=\frac{1}{k^2}\int\limits_M dx\;g^{1/2}
\left(R-2\Lambda\right)\;,
\ee
where $R$ is the scalar curvature, $k^2=16\pi G$ is the Einstein
coupling constant, $G$ is the Newtonian gravitational
constant and $\Lambda$ is the cosmological constant.
Here we neglect the boundary term for simplicity; it will
not affect our calculations.

The first variation of the action gives the 
classical equations of motion
\be
g^{-1/2}{\delta S\over \delta g_{\mu\nu}}
=-\frac{1}{k^2}\left(R^{\mu\nu}-{1\over 2}g^{\mu\nu}R
+\Lambda g^{\mu\nu}\right)\,,
\ee
which satisfy, of course, the N\"other identities
\be
\nabla_\mu \left(R^{\mu\nu}-{1\over 2}g^{\mu\nu}R
+\Lambda g^{\mu\nu}\right)=0\,.
\ee
Here $R_{\mu\nu}$ is the Ricci tensor defined in terms of the Riemann
tensor by
$R_{\mu\nu}=R^\alpha{}_{\mu\alpha\nu}$.

The second variation of the action defines a second-order
partial differential operator by
\be
g^{-1/2}\frac{\delta^2 S}{\delta g_{\mu\nu}\delta g_{\alpha\beta}}
h_{\alpha\beta}
=P^{\mu\nu\alpha\beta}h_{\alpha\beta}\,,
\ee
where
\bea
P^{\mu\nu,\alpha\beta}
&=&
-{1\over 2 k^2}
\Biggl\{
-\left(g^{\alpha(\mu}g^{\nu)\beta}
-g^{\alpha\beta}g^{\mu\nu}\right)\Delta
\nonumber\\[10pt]
&&
-g^{\mu\nu}\nabla^{(\alpha}\nabla^{\beta)}
-g^{\alpha\beta}\nabla^{(\mu}\nabla^{\nu)}
+2\nabla^{(\mu}g^{\nu)(\alpha}\nabla^{\beta)}
\nonumber\\[10pt]
&&
-2R^{(\mu|\alpha|\nu)\beta}
-g^{\alpha(\mu}R^{\nu)\beta}
-g^{\beta(\mu}R^{\nu)\alpha}
+R^{\mu\nu}g^{\alpha\beta}
+ R^{\alpha\beta}g^{\mu\nu}
\nonumber\\
&&
+\left(g^{\mu(\alpha}g^{\beta)\nu}
-{1\over 2}g^{\mu\nu}g^{\alpha\beta}\right)
(R-2\Lambda)
\Biggr\}
\eea
Here, of course, $\Delta=g^{\mu\nu}\nabla_\mu\nabla_\nu$
denotes the Laplacian.

Next, we choose 
the De~Witt gauge condition
\be
\chi^\alpha
=-E^{\alpha\beta\mu\nu}\nabla_\beta h_{\mu\nu}
=-\left(g^{\alpha(\nu}\nabla^{\mu)}
-\varkappa g^{\mu\nu}\nabla^\alpha \right)
h_{\mu\nu}\;.
\ee
The ghost operator in this gauge is a second-order
differential operator defined by 
\be
F^{\mu}{}_{\nu}
=-2E^{\mu\alpha\beta}{}_\nu\nabla_\alpha\nabla_\beta
=-\delta^{\mu}_{\nu}\Delta
+(2\varkappa-1)\nabla^\mu\nabla_\nu
-R^{\mu}_{\nu}\;.
\ee
For this operator to be non-singular, the gauge parameter should satisfy
the condition $\varkappa\ne 1$.

For the graviton operator $L$ to be
non-degenerate it is necessary to choose
the operator $\gamma$ as a zero order
differential operator defined by
\be
\gamma_{\mu\nu}
=\frac{\alpha}{k^2} g_{\mu\nu}\;,
\ee
where $\alpha\ne 0$ is a real parameter. 
Thus we obtain a two-parameter class of gauges involving
two arbitrary parameters, $\varkappa$ and $\alpha$.

The graviton operator $L$ now reads
\bea
L^{\mu\nu,\alpha\beta}
&=&
{1\over 2 k^2}
\Biggl\{
-\left(g^{\alpha(\mu}g^{\nu)\beta}
-(1+2\alpha\varkappa^2)
g^{\alpha\beta}g^{\mu\nu}\right)\Delta
\nonumber\\[10pt]
&&
-(1+2\alpha\varkappa)g^{\mu\nu}\nabla^{(\alpha}\nabla^{\beta)}
-(1+2\alpha\varkappa)g^{\alpha\beta}\nabla^{(\mu}\nabla^{\nu)}
+2(1+\alpha)\nabla^{(\mu}g^{\nu)(\alpha}\nabla^{\beta)}
\nonumber\\[10pt]
&&
-2R^{(\mu|\alpha|\nu)\beta}
-g^{\alpha(\mu}R^{\nu)\beta}
-g^{\beta(\mu}R^{\nu)\alpha)}
+R^{\mu\nu}g^{\alpha\beta}
+g^{\mu\nu}R^{\alpha\beta}
\nonumber\\
&&
+\left(g^{\mu(\alpha}g^{\beta)\nu}
-{1\over 2}g^{\mu\nu}g^{\alpha\beta}\right)
(R-2\Lambda)
\Biggr\}
\eea
The most convenient choice is the so-called minimal gauge
\be
\varkappa=\frac{1}{2}\,, \qquad
\alpha=-1
\,.
\ee
In this gauge the non-diagonal derivatives in both the graviton
operator and the ghost operator vanish
\bea
L^{\mu\nu,\alpha\beta}
&=&
{1\over 2 k^2}
\Biggl\{
\left(g^{\alpha(\mu}g^{\nu)\beta}
-\frac{1}{2}
g^{\alpha\beta}g^{\mu\nu}\right)(
-\Delta+R-2\Lambda)
\nonumber\\[10pt]
&&
-2R^{(\mu|\alpha|\nu)\beta}
-g^{\alpha(\mu}R^{\nu)\beta}
-g^{\beta(\mu}R^{\nu)\alpha)}
+R^{\mu\nu}g^{\alpha\beta}
+g^{\mu\nu}R^{\alpha\beta}
\Biggr\}\,,
\eea
\be
F^{\mu}{}_{\nu}
=-\delta^{\mu}{}_{\nu}\Delta
-R^{\mu}{}_{\nu}\;.
\ee 

Finally, we 
define the graviton operator in the canonical Laplace-type form,
$\hat L$, by factoring out the configuration space metric
(in the minimal gauge $\varkappa=1/2$)
\be
\hat L_{\mu\nu}{}^{\alpha\beta}=2k^2 E^{-1}_{\mu\nu\rho\sigma}
L^{\rho\sigma\alpha\beta}\,.
\ee
We obtain
\be
\hat L_{\mu\nu}{}^{\alpha\beta}
=-\delta^{\alpha}_{(\mu}\delta^{\beta}_{\nu)}\Delta
+Q_{\mu\nu}{}^{\alpha\beta}\,,
\ee
where
\bea
Q_{\mu\nu}{}^{\alpha\beta}
&=&
-2R_\mu{}^{(\alpha}{}_\nu{}^{\beta)}
-2\delta^{(\alpha}_{(\mu}R_{\nu)}^{\beta)}
+R_{\mu\nu}g^{\alpha\beta}
+\frac{2}{n-2}g_{\mu\nu}R^{\alpha\beta}
\nonumber\\
&&
+\left(\delta^\alpha_{(\mu}\delta^\beta_{\nu)}
-\frac{1}{(n-2)}g_{\mu\nu}g^{\alpha\beta}\right)R
-2\Lambda\delta^\alpha_{(\mu}\delta^\beta_{\nu)}
\,.
\eea

One can show that the contribution of the determinant of the
operator $\gamma$ can be neglected (more precisely, it can be
absorbed in the measure of the path integral) 
since it is of zero order.
Thus, with this choice of gauge parameters the one-loop
effective action
of quantum general relativity is given by
\be
\Gamma_{(1)}=-\frac{1}{2i}\log\Det\hat L
+\frac{1}{i}\log\Det F
\,.
\label{eaxxz}
\ee
Therefore, in order to compute the effective action we need to
compute the determinants of Laplace type partial differential
operators acting on symmetric two-tensors and vectors.


\section{Heat Kernel Method}
\setcounter{equation}0

As we described in the previous section
the effective action in quantum field
theory can be computed within the semi-classical perturbation theory. 
It is determined by the functional determinants of
second-order hyperbolic partial differential operators with Feynman boundary
conditions and the higher-loop approximations are determined in terms of the
Feynman propagators and the classical vertex functions. As we noted above these
expressions are purely formal and need to be regularized and renormalized,
which can be done in a consistent way in renormalizable field theories. One
should stress, of course, that many physically interesting theories (including
Einstein's general relativity) are 
perturbatively non-renormalizable. Since we only need
Feynman propagators we can do the Wick rotation and consider instead of
hyperbolic operators the elliptic ones. The Green functions of elliptic
operators and their functional determinants can be expressed in terms of
the heat kernel. That is why we concentrate below on the
calculation of the heat kernel.

The heat kernel is one of the most powerful tools in mathematical
physics and  geometric analysis (see, for example, the books 
\cite{gilkey95,berline92,avramidi00,kirsten01} and reviews 
\cite{camporesi90,avramidi99a,avramidi02,vassilevich03,barvinsky85}).
The  short-time  asymptotic expansion of the trace of the heat kernel
determines the spectral asymptotics of the differential operator. The
coefficients of this asymptotic expansion, called the  heat invariants,
are extensively used in geometric analysis, in particular, in spectral
geometry and index theorems proofs
\cite{gilkey95,berline92}.

The gauge invariance (or covariance) in  quantum gauge field theory and quantum
gravity is of fundamental importance. That is why, 
manifestly covariant methods
present inestimable advantage. A manifestly
covariant calculus is such that every step is expressed in terms of geometric
objects; it does not have some intermediate non-covariant steps that lead
to an invariant result. Below we describe a manifestly covariant method
for calculation of the heat kernel following mainly our papers
\cite{avramidi90b,avramidi91,avramidi99a,avramidi00,avramidi02}. 

\subsection{Laplace Type Operators}

Let $(M,g)$  be a smooth compact Riemannian manifold of dimension $n$ without
boundary, equipped with a positive definite Riemannian metric
$g$. 
We assume that it is complete simply connected orientable and spin.
We
denote the local coordinates on $M$ by $x^\mu$, with Greek indices
running over $1,\dots, n$. Let $e_{a}{}^\mu$ be a local orthonormal
frame defining a basis for the tangent space $T_xM$.
We denote the frame indices by low case Latin indices from the beginning
of the alphabet, which also run over $1,\dots,n$. The frame indices are
raised and lowered by the metric $\delta_{ab}$.  Let $e^{a}{}_\mu$ be the
matrix inverse to $e_{a}{}^\mu$, defining the dual basis in the
cotangent space $T_x^*M$.
As usual, the orthonormal frame,
$e^a{}_\mu$ and $e_a{}^\mu$, 
will be used to transform the coordinate
(Greek) indices to the orthonormal (Latin) indices. 
The Riemannian volume element is defined as usual by
$
d\vol=dx\,g^{1/2}\,,
$
where
$
g=\det g_{\mu\nu}=(\det e_a{}^\mu)^2\,.
$
The spin connection $\omega^{ab}{}_{\mu}$ is defined in terms of
the covariant derivatives of the orthonormal frame with the Levi-Civita
connection.
The curvature of the spin connection defines the Riemann tensor,
$R^{a}{}_{b\mu\nu}$, the Ricci
tensor, $R_{\mu\nu}=R^\alpha{}_{\mu\alpha\nu}$,  and the scalar curvature, 
$R=R^\mu{}_\mu$, as usual.

Let $\mathcal{T}$ be a spin-tensor bundle realizing a representation
$\Sigma$  of the spin group \mbox{$\Spin(n)$}, the double covering of
the group $SO(n)$, with the fiber $\Lambda$. 
Let $\Sigma_{ab}$ be the generators of the 
orthogonal algebra $\mathcal{SO}(n)$, the Lie algebra of the orthogonal
group $SO(n)$.
The spin connection induces a connection  on the bundle $\mathcal{T}$
defining the covariant derivative of smooth sections  $\varphi$ of the
bundle $\mathcal{T}$ by
\be
\nabla^{\rm spin}_\mu\varphi
=\left(\partial_\mu
+\frac{1}{2}\omega^{ab}{}_{\mu}\Sigma_{ab}\right)\varphi\,.
\ee
The commutator of covariant derivatives
defines the curvature of this connection via
\be
[\nabla^{\rm spin}_\mu,\nabla^{\rm spin}_\nu]\varphi
=\frac{1}{2}R^{ab}{}_{\mu\nu}\Sigma_{ab}\varphi\,.
\ee
The covariant derivative  along the frame vectors is 
defined by
$\nabla_a=e_a{}^\mu\nabla_\mu$.
For example, with our notation, $\nabla_a\nabla_b T_{cd}
=e_a{}^\mu e_b{}^\nu e_c{}^\alpha e_d{}^\beta\nabla_\mu\nabla_\nu 
T_{\alpha\beta}$.
The metric $\delta_{ab}$ induces a positive definite fiber metric on
tensor bundles.  


Let $G_{YM}$ be a  compact Lie group (called a gauge group). It
naturally defines the principal fiber bundle over the manifold $M$  with
the structure group $G_{YM}$. We consider a representation of the
structure group $G_{YM}$ and the associated vector bundle through
this representation with the same structure group $G_{YM}$ whose typical
fiber is a $k$-dimensional  vector space $W$. Then for any spin-tensor
bundle $\mathcal{T}$ we define the 
{\it twisted spin-tensor bundle} $\mathcal{V}$
via the twisted product of the bundles $\mathcal{W}$ and $\mathcal{T}$. The
fiber of the bundle $\mathcal{V}$ is $V=\Lambda\otimes W$ so that the
sections of the bundle $\mathcal{V}$ are represented locally by $k$-tuples
of spin-tensors.

Let $\mathcal{A}^{YM}$ be a connection one form on the bundle
$\mathcal{W}$
(called Yang-Mills or gauge connection) taking values in the Lie
algebra $\mathcal{G}_{YM}$ of the gauge group $G_{YM}$. Then the total
connection on the bundle $\mathcal{V}$ is defined by
\be
\nabla_\mu\varphi
=\left(\partial_\mu+{\cal A}_\mu\right)\varphi\,,
\ee
where
\be
{\cal A}_\mu=
\frac{1}{2}\omega^{ab}{}_\mu \Sigma_{ab}\otimes \II_W
+\II_\Lambda\otimes\mathcal{A}^{YM}_\mu\,,
\ee
and the total curvature ${\cal R}$ of the bundle $\mathcal{V}$ is
defined by
\be
[\nabla_\mu,\nabla_\nu]\varphi
={\cal R}_{\mu\nu}\varphi\,,
\label{219mm}
\ee
where
\be
{\cal R}_{\mu\nu}=
\frac{1}{2}R^{ab}{}_{\mu\nu} \Sigma_{ab}
+{\cal R}^{YM}_{\mu\nu}\,,
\label{220}
\ee
and
\be
\mathcal{R}^{YM}_{\mu\nu}
=\partial_\mu\mathcal{A}^{YM}_\nu
-\partial_\nu\mathcal{A}^{YM}_\mu
+[\mathcal{A}^{YM}_\mu,\mathcal{A}^{YM}_\mu]
\ee
is the curvature of the Yang-Mills connection.

We also consider the bundle $\End({\cal V})$
of endomorphisms of the bundle $\mathcal{V}$.
The covariant derivative  of sections
of this bundle is defined by
\be
\nabla_\mu Q
=\partial_\mu Q+[\mathcal{A}_\mu,Q]\,,
\ee
and the commutator of covariant derivatives is equal to
\be
[\nabla_\mu,\nabla_\nu]Q
=[\mathcal{R}_{\mu\nu},Q]\,.
\label{225}
\ee

We
assume that the vector bundle ${\cal V}$ is equipped with a Hermitian metric. 
This
naturally identifies the dual vector bundle ${\cal V}^*$ with ${\cal V}$.
We assume that the connection $\nabla$ is
compatible with the Hermitian metric on the vector bundle ${\cal V}$. 
The
connection is given its unique natural extension to bundles in the tensor
algebra over ${\cal V}$ and ${\cal V}^*$.  In fact, using the Levi-Civita
connection
of the metric $g$ together with the connection on the bundle
${\cal V}$, we naturally
obtain connections on all bundles in the tensor algebra over
${\cal V},\,{\cal V}^*,\,TM$
and $T^*M$; the resulting connection will usually be denoted just by $\nabla$.
It is usually clear which bundle's connection is being referred to, from the
nature of the section being acted upon.

We denote by $C^\infty({\cal V})$ the space of smooth sections of the bundle
${\cal V}$. The fiber inner product on the bundle ${\cal V}$ 
defines a
natural $L^2$ inner product and the  $L^2$-trace $\Tr$ using the
invariant Riemannian measure on the manifold $M$.  
The completion of
$C^\infty({\cal V})$ in this norm defines the Hilbert space $L^2({\cal V})$ of
square
integrable sections. 
Let $\nabla^*$ be the formal adjoint
to $\nabla$ defined using the Riemannian metric and
the Hermitian structure  on ${\cal V}$ and let $Q$
be a smooth
Hermitian section of the endomorphism bundle $\End({\cal V})$. 

A Laplace type
operator $L: C^\infty(V)\to C^\infty(V)$ is a partial differential operator
of the form  
\be 
L=\nabla^*\nabla+Q=-\Delta+Q\,.
\label{1ms}
\ee
In local coordinates the Laplacian is defined by
\be
\Delta=g^{\mu\nu}\nabla_\mu\nabla_\nu=
g^{-1/2}(\partial_\mu+{\cal A}_\mu)g^{1/2}g^{\mu\nu}
(\partial_\nu+{\cal A}_\nu).
\ee
and, therefore, 
\bea
L&=&
-g^{-1/2}(\partial_\mu+{\cal A}_\mu)g^{1/2}g^{\mu\nu}
(\partial_\nu+{\cal A}_\nu)+Q
\nonumber\\
&=&
-g^{\mu\nu}\partial_\mu\partial_\nu-2a^\mu\partial_\mu+q,
\eea
where 
\be
a^\mu=g^{\mu\nu}{\cal A}_\nu
+{1\over 2}g^{-1/2}\partial_\nu(g^{1/2}g^{\nu\mu})
\ee
\be
q=Q-g^{\mu\nu}{\cal A}_\mu{\cal A}_\nu
-g^{-1/2}\partial_\mu(g^{1/2}g^{\mu\nu}{\cal A}_\nu).
\ee

Thus, a Laplace type operator is constructed from the following three pieces of
geometric data: i) a Riemannian metric $g$ on $M$, which determines the
second-order part, ii) a connection $1$-form ${\cal A}$ on the vector bundle
${\cal V}$, which determines the first-order part, iii) an endomorphism $Q$ of
the vector bundle ${\cal V}$, which determines the zeroth order part. It is
worth noting that every second-order differential operator with a scalar
leading symbol given by the metric tensor is of Laplace type and can be put in
this form by choosing the appropriate connection and the endomorphism $Q$.


It is easy to show that the Laplacian, $\Delta$, and, therefore, the operator
$L$, is an elliptic symmetric partial differential operator satisfying 
\be
(L\varphi,\psi)=(\varphi,L\psi), 
\ee
with a positive principal symbol.
Moreover, the operator $L$ is essentially self-adjoint, i.e., it has a unique
self-adjoint extension. We will not be very careful about distinguishing
between the operator $L$ and its closure, and will simply say that the operator
$L$ is elliptic and self-adjoint.

It is well
known \cite{gilkey95} that: 
\begin{itemize}
\item[i)]
the operator $L$ has a discrete real spectrum,
$\{\lambda_n\}_{n=1}^\infty$, bounded from below: 
\be
\lambda_0< \lambda_1 < \lambda_2 < \cdots<\lambda_n<\cdots
\ee
with some real constant $\lambda_0$, 
\item[ii)]
the eigenvalues grow as $k\to \infty$ as $\lambda_k\sim Ck^{2/n}$,
where $n=\dim M$, 
\item[iii)]
all
eigenspaces of the operator $L$ are finite-dimensional, and 
\item[iv)]
the
eigenvectors, $\{\varphi_n\}_{n=1}^\infty$, of the operator $L$, 
are smooth
sections of the vector bundle ${\cal V}$ that form a complete orthonormal basis
in $L^2({\cal V})$.
\end{itemize}

\subsection{Spectral Functions}

The spectrum of the operator $L$ can be described by cerain
spectral invariants, called spectral
functions. First of all, we define the {\it heat trace}:
\be
\Theta(t)=\sum_{n=1}^\infty e^{-t\lambda_n}\,,
\ee
where each eigenvalue is counted with multiplicities.
The heat trace 
is well defined for real positive $t$. Notice that it can
be
analytically continued to an analytic function of $t$
in the right half-plane (for ${\rm Re}\, t>0$).

The heat trace 
determines other spectral functions by integral transforms:
the {\it distribution function} 
(also called counting function), 
defined as the number of
eigenvalues below the level $\lambda$,
\be
N(\lambda)=\sum_{n=1}^\infty\theta(\lambda-\lambda_n)
=\frac{1}{2\pi i}
\int\limits_{\varepsilon-i\infty}^{\varepsilon+i\infty}\frac{dt}{t}\;
e^{t\lambda}\,\Theta(t),
\ee
where $\varepsilon$ is a positive constant,
the {\it density function}, 
\be
\rho(\lambda)=\sum_{n=1}^\infty \delta(\lambda-\lambda_n)
=\frac{1}{2\pi i}
\int\limits_{\varepsilon-i\infty}^{\varepsilon+i\infty}dt\;
e^{t\lambda}\,\Theta(t),
\ee
and the {\it zeta-function},
\be
\zeta(s,\lambda)=\sum_{n=1}^\infty \frac{1}{(\lambda_n-\lambda)^s}
={1\over\Gamma(s)}\int\limits_0^\infty dt\; t^{s-1}\,
e^{t\lambda}\Theta(t),
\ee
where $\lambda$ is a large negative constant such that
${\rm Re}\,\lambda<\lambda_0$ and $s$ is a complex parameter with 
${\rm Re}\, s>n/2$.

In principle, if known {\it exactly}, they determine the spectrum.
Of course, this is not valid for asymptotic expansions of the spectral
functions.
There are examples of operators that have the same asymptotic series of the
spectral
functions but different spectrum.

The zeta function enables one to define, in particular, the 
{\it zeta-regularized
determinant} of the operator 
$(L-\lambda)$,
\be
\zeta'(0,\lambda)\equiv {\partial\over\partial s}\zeta(s,\lambda)\Big|_{s=0}
=-\log\Det (L-\lambda),
\ee
which determines the one-loop effective action in quantum field theory.

\subsection{Heat Kernel}

{}For $t>0$ the operators 
\be
U(t)=\exp(-tL)
\ee
form a semi-group of bounded
operators on $L^2({\cal V})$, so called heat semi-group. The kernel of this
operator
is defined by  
\be 
U(t|x,x')
=\sum\limits_{n=1}^\infty
e^{-t\lambda_n}\varphi_n(x)\otimes \varphi^*_n(x'),
\ee
where each eigenvalue is counted with multiplicities.
It is a section of the external tensor product of vector bundles 
${\cal V}\boxtimes
{\cal V}^*$ over $M\times M$, which 
can also be regarded as an endomorphism from the fiber of ${\cal V}$ 
over $x'$
to the
fiber of ${\cal V}$ over $x$. This kernel satisfies the heat equation 
\be
\left(\partial_t+L\right)U(t)=0
\label{he-5/01}
\ee
with the initial condition
\be
U(0^+|x,x')=\delta(x,x')\,
\label{init-5/01}
\ee
and is called the {\it heat kernel}.

Moreover, the heat semigroup $U(t)$ is a trace-class operator  with a well
defined $L^2$-trace, 
\be
\Tr\exp(-tL)=\int\limits_M d\vol\; \tr_V U^{\rm diag}(t)\,.
\ee
Hereafter
$\tr_V$ denotes the fiber trace and the label `${\rm diag}$' means the diagonal
value of a two-point quantity, e.g.  
\be
U^{\rm diag}(t|x)=U(t|x,x')\Big|_{x=x'}\,.
\ee
It is easy to see that the heat trace defined above is equal to the trace of
the heat semigroup, that is,
\be
\Theta(t)=\Tr\exp(-tL)\,.
\ee


\subsection{Asymptotic Expansion of the Heat Kernel}

In the following we are going to study the heat kernel only locally, i.e.
in the neighbourhood of the diagonal of $M\times M$, when the points $x$ and
$x'$ are
close to each other. The exposition will follow mainly
our papers \cite{avramidi91,avramidi00,avramidi99a,avramidi02}.
We will keep a point $x'$ of the manifold fixed and consider a
small geodesic ball, i.e. a small neighbourhood of the point $x'$:
$B_\varepsilon(x')
=\{x\in M| r(x,x')<\varepsilon\}$, $r(x,x')$ being the
geodesic distance
between the
points $x$ and $x'$. We will take the radius of the ball sufficiently small, so
that each
point $x$ of the ball
of this neighbourhood can be connected by a unique geodesic
with the point $x'$. This can be always
done if the size of the ball is smaller than the injectivity radius of the
manifold,
$\varepsilon<r_{\rm inj}$.

Let $\sigma(x,x')$ be the geodetic interval, also called world function,
defined as one half the square of the length
of the geodesic connecting the points $x$ and $x'$
\be
\sigma(x,x')={1\over 2}r^2(x,x').
\ee
The first derivatives of this function with respect to $x$ and $x'$
define tangent vector fields to the geodesic at the points
$x$ and $x'$ respectively
pointing in opposite directions
\bea
u^\mu &=& g^{\mu\nu}\nabla_\nu\sigma,
\\
u^{\mu'} &=& g^{\mu'\nu'}\nabla'_{\nu'}\sigma,
\label{99xxx}
\eea
and the determinant of the mixed second derivatives defines a so called
Van Vleck-Morette determinant
\be
\Delta(x,x')=g^{-1/2}(x)
\det\left[-\nabla_\mu\nabla'_{\nu'}\sigma(x,x')\right]g^{-1/2}(x').
\ee
This object should not be confused with the Laplacian, which is also denoted by
$\Delta$. 

Let, finally, ${\cal P}(x,x')$ denote the parallel transport operator
of sections of the vector bundle ${\cal V}$ along the geodesic from the point
$x'$ to the point $x$. It is a section of the external tensor product of the
vector bundle ${\cal V}\boxtimes{\cal V}^*$ over $M\times M$, or, in other
words, it is an endomorphism from the fiber of ${\cal V}$ over $x'$ to the
fiber of ${\cal V}$ over $x$. Here and everywhere below the coordinate indices
of the tangent space at the point $x'$ are denoted by primed Greek letters.
They are raised and lowered by the metric tensor $g_{\mu'\nu'}(x')$ at the
point $x'$.  The derivatives with respect to $x'$ will be denoted by primed
Greek indices as well. 


We
extend the local orthonormal frame $e_a{}^{\mu'}(x')$ at the point $x'$ to
a local  orthonormal frame $e_a{}^\mu(x)$ at the point $x$ by parallel
transport.
The parameters of the geodesic connecting the points $x$ and $x'$,
namely the unit tangent vector at the point $x'$  and the length of the
geodesic, (or, equivalently, the tangent vector at the point $x'$ with
the norm equal to the length of the geodesic), provide normal coordinate
system for $B_\varepsilon(x')$. 
Now, let us define the following geometric parameters
\be
y^a
=e^{a}{}_{\mu}u^{\mu}
=-e^{a}{}_{\mu'}u^{\mu'}\,,
\label{ncxxz}
\ee
so that
\be
u^{\mu}=e_{a}{}^{\mu}y^a
\qquad
\mbox{and}
\qquad
u^{\mu'}=-e_{a}{}^{\mu'}y^a
\,.
\ee
Notice that $y^a=0$ at $x=x'$.
The geometric parameters $y^a$ are nothing but the normal coordinates.


Near the diagonal of $M\times M$ all these two-point functions are smooth
single-valued
functions of the coordinates of the points $x$ and $x'$.
Let us note from the beginning that we will construct the heat kernel in form
of
covariant Taylor series in coordinates. In the smooth case these series do not
necessarily
converge. However, if one assumes additionally that the two-point funtions are
analytic,
then the Taylor series converge   in a sufficiently small neighborhood of the
diagonal.

Further, one can easily prove that the function
\be
U_0(t|x,x')=(4\pi t)^{-n/2}\Delta^{1/2}(x,x')\exp\left(-{1\over
2t}\sigma(x,x')\right){\cal P}(x,x')
\ee
satisfies the initial condition
\be
U_0(0^+|x,x')=\delta(x,x').
\ee
Moreover, locally it also satisfies
the heat equation in the free case, when the Riemannian
curvature of the manifold, ${\rm Riem}$,
the curvature of the bundle connection, ${\cal R}$, and the endomorphism
$Q$ vanish:
\be
{\rm Riem}={\cal R}=Q=0\,.
\ee
Therefore, $U_0(t|x,x')$ is the exact heat kernel for a pure 
Laplacian
in flat Euclidean space with a flat
trivial bundle connection and without the endomorphism $Q$.

\subsubsection{Transport Function}

This function gives a good framework for the approximate solution in the
general case.
Namely, by factorizing out this free factor we get an ansatz
\be
U(t|x,x')=(4\pi t)^{-n/2}\Delta^{1/2}(x,x')\exp\left(-{1\over
2t}\sigma(x,x')\right){\cal P}(x,x')
\Omega(t|x,x').
\label{150}
\ee
The function $\Omega(t|x,x')$, called the {\it transport function}, is a
section of the
endomorphism vector bundle $\End(V)$ over the point $x'$.
Using the definition of the functions $\sigma(x,x')$, $\Delta(x,x')$ and ${\cal
P}(x,x')$
it is not difficult to find that the transport function satisfies a transport
equation
\be
\left(\partial_t+{1\over t}D+\tilde L\right)\Omega(t)=0,
\ee
where $D$ is the radial vector field, i.e. operator of differentiation along
the geodesic,
defined by
\be
D=u^\mu\nabla_\mu,
\ee
and $\tilde L$ is a second-order differential operator defined by
\be
\tilde L={\cal P}^{-1}\Delta^{-1/2}L\Delta^{1/2}{\cal P}.
\label{160}
\ee
The initial condition for the transport function is obviously
\be
\Omega(t|x,x')=\II_V,
\ee
where $\II_V$ is the identity endomorphism of the vector bundle ${\cal V}$ over
$x'$.

It is obvious that if we replace the operator $L$ by
$(L-\lambda)$, with ${\rm Re}\, \lambda<\lambda_0$, then the heat kernel and
the transport function are simply multiplied by $e^{t\lambda}$, i.e.
the transport function for the operator $(L-\lambda)$ is
$e^{t\lambda}\Omega(t)$.
Further, for $\lambda<\lambda_0$ the operator $(L-\lambda)$
becomes a positive operator. Therefore, the function
$e^{t\lambda}\Omega(t)$
satisfies the following asymptotic conditions
\be
\lim_{t\to\infty,0}t^\alpha\partial_t^N\left[e^{t\lambda}\Omega(t)\right]=0
\qquad
{\rm for\ }\ \lambda<\lambda_1, \ \alpha>0, \ N\ge 0.
\label{200}
\ee
In other words, as $t\to\infty$ the function $e^{t\lambda}\Omega(t)$ and all
its derivatives
decreases faster than any power of $t$, actually it decreases exponentialy, and
as $t\to 0$ the product of $e^{t\lambda}\Omega(t)$ with any positive power of
$t$ vanishes.

Hereafter we fix $\lambda<\lambda_0$, so that $(L-\lambda)$ is a positive
operator.
Now, let us consider a slightly modified version of the Mellin transform of the
function $e^{t\lambda}\Omega(t)$ introduced in \cite{avramidi91}
\be
b_q(\lambda)={1\over \Gamma(-q)}
\int\limits_0^\infty dt\;
t^{-q-1}e^{t\lambda}\Omega(t).
\label{41-7/98}
\ee
Note that for fixed $\lambda$ this is a Mellin transform of
$e^{t\lambda}\Omega(t)$
and for a fixed $q$ this is a Laplace transform of the function
$t^{-q-1}\Omega(t)$.
The integral (\ref{41-7/98}) converges for ${\rm Re}\, q<0$.
By integrating by parts $N$ times and using the asymptotic conditions
(\ref{200})
we also get
\be
b_q(\lambda)
={1\over \Gamma(-q+N)}\int\limits_0^\infty dt\;
t^{-q-1+N}(-\partial_t)^N
\left[e^{t\lambda}\Omega(t)\right].
\ee
This integral converges for ${\rm Re}\,q<N-1$.
Using this representation one can prove that \cite{avramidi91}
the function $b_q(\lambda)$ is an entire function of $q$ (analytic 
everywhere) satisfying the 
the asymptotic condition
\be
\lim_{|q|\to\infty,\ {\rm Re}\,q<N}\Gamma(-q+N)b_q(\lambda)=0,\qquad
{\rm for\ any}\ N>0.
\label{350}
\ee
Moreover, 
the values of the function $b_q(\lambda)$ at the integer positive points
$q=k$ are given by
\be
b_k(\lambda)=(-\partial_t)^k\left[e^{t\lambda}\Omega(t)\right]\Big|_{t=0}
=\sum_{n=0}^k{k\choose n} a_{n}\,,
\label{43-7/98}
\ee
where
\be
a_k=(-\partial_t)^k\Omega(t)\Big|_{t=0},
\ee

By inverting the Mellin transform we obtain a new ansatz for the transport
function
and, hence, for the heat kernel
\be
\Omega(t)={1\over 2\pi i}\int\limits_{c-i\infty}^{c+i\infty}dq\;
e^{-t\lambda}t^q\,\Gamma(-q)b_q(\lambda)
\label{46-7/98}
\ee
where $c<0$ and ${\rm Re}\,\lambda<\lambda_0$.
Clearly, since the left-hand side of this equation does not depend on
$\lambda$,
neither does the right hand side. Thus, $\lambda$ serves as an auxiliary
parameter that
regularizes the behavior at $t\to \infty$.
If we invert instead the Laplace transform, we obtain another representation
\be
\Omega(t)=
{1\over 2\pi
i}\int\limits_{\gamma-i\infty}^{\gamma+i\infty}d\lambda\;
e^{-t\lambda}t^{q+1}\Gamma(-q)b_q(\lambda)
\ee
where $\gamma<\lambda_0$ and ${\rm Re}\,q<0$.

Substituting this ansatz into the transport equation we get a functional
equation for
the function $b_q$
\be
\left(1+{1\over q}D\right)b_q(\lambda)
=(\tilde L-\lambda)\,b_{q-1}(\lambda).
\label{400}
\ee
The initial condition for the transport function is translated into
\be
b_0(\lambda)=\II_V.
\label{500}
\ee

Thus, we have reduced the problem of solving the heat equation to the following
problem:
one has to find an entire function of $q$, 
$b_q(\lambda|x,x')$, that satisfies
the
functional equation
(\ref{400}) with the initial condition (\ref{500}) and the asymptotic condition
(\ref{350}).

Although the variables $q$ and $\lambda$ seem to be independent they are very
closely
related to each other. In particular, by differentiating with respect to
$\lambda$ we obtain
an important result
\be
{\partial\over \partial \lambda}b_q(\lambda)=-qb_{q-1}(\lambda).
\ee
Also, by differentiating the eq. (\ref{400}) with respect
to $q$ one obtains another recursion 
\be
\left(1+{1\over q}D\right)b'_{q}(\lambda)
=\tilde L\,b'_{q-1}(\lambda)+{1\over q^2}Db_{q}(\lambda),
\label{12b}
\ee
where
\be
b'_{q}(\lambda)={\partial\over\partial q}b_{q}(\lambda),
\ee
which enables one to compute the derivatives of the function $b_{q}(\lambda)$
at positive integer points if one fixes its value $b'_{0}(\lambda)$.
This tirns out to be useful when computing the determinant of the operator 
$(L-\lambda)$.

Moreover,  one can actually manifest the dependence of $b_q(\lambda)$
on $\lambda$. 
It is not difficult to prove that \cite{avramidi91} the integral
\be
b_q(\lambda)={1\over 2\pi i}\int\limits_{c_1-i\infty}^{c_1+i\infty}dp\,
{\Gamma(-p)\Gamma(p-q)\over\Gamma(-q)}(-\lambda)^{q-p}a_p,
\label{122xxz}
\ee
with ${\rm Re}\,q<c_1<0$,
satisfies the equation (\ref{400}) if $a_p$ satisfies this equation for
$\lambda=0$, i.e.
\be
\left(1+{1\over q}D\right)a_q=\tilde L\,a_{q-1}.
\label{400-7/98}
\ee
with the initial condition
\be
a_0=\II_V.
\label{500-7/98}
\ee
{}For integer $q=k=1,2,\dots$ the functional equation (\ref{400-7/98}) becomes
a recursion system
that, together with the initial condition (\ref{500-7/98}),
determines all coefficients $a_k$.

Now, from eq. (\ref{122xxz})
we also obtain the asymptotic expansion of $b_q(\lambda)$ as
$\lambda\to -\infty$,
\be
b_q(\lambda)
\sim\sum_{n=0}^\infty
{\Gamma(q+1)\over n!\Gamma(q-n+1)}(-\lambda)^{q-n}a_{n}.
\ee
For integer $q$ this coincides with (\ref{43-7/98}).

The function $b_q(\lambda)$ turns out to be extremely useful in computing the
heat kernel, the resolvent
kernel, the zeta-function and the determinant of the operator $L$. 
It contains
the same
information about the operator $L$ as the heat kernel.
In some cases the function $b_q(\lambda)$ 
can be constructed just by analytical
continuation from
the integer positive values $b_k$ \cite{avramidi91}.

\subsubsection{Asymptotic Expansion of the Transport Function}

Now we are going to do the usual trick, namely, to move the contour of
integration over $q$
to the right. Due to the presence of the gamma function $\Gamma(-q)$ the
integrand
has simple poles at the non-negative integer
points $q=0,1,2\dots$, which contribute to the integral while moving the
contour.
So, we get
\be
\Omega(t)=e^{-t\lambda}
\left\{\sum\limits_{k=0}^{N-1}{(-t)^k\over k!}b_k(\lambda)+R_N(t)
\right\},
\label{300}
\ee
where
\be
R_N(t)={1\over 2\pi i}
\int\limits_{c_N-i\infty}^{c_N+i\infty}dq\,t^q\,\Gamma(-q)
b_q(\lambda)
\ee
with $c_N$ is a constant satisfying the condition $N-1<c_N<N$.
As $t\to 0$ the rest term $R_N(t)$ behaves like $O(t^N)$, so we obtain an
asymptotic expansion as $t\to 0$
\be
\Omega(t)\sim e^{-t\lambda}
\sum\limits_{k=0}^\infty{(-t)^k\over k!}b_k(\lambda)
=\sum\limits_{k=0}^\infty{(-t)^k\over k!}a_k.
\label{600}
\ee

Using our ansatz (\ref{150}) we find immediately the heat trace
\be
\Theta(t)=(4\pi t)^{-n/2}e^{-t\lambda}
{1\over 2\pi i}
\int\limits_{c-i\infty}^{c+i\infty}dq\,t^q\,\Gamma(-q)B_q(\lambda),
\ee
where
\be
B_q(\lambda)=\int\limits_M d\vol\;\tr_V\,b^{\rm diag}_q(\lambda).
\ee
The heat trace has an analogous asymptotic expansion as
$t\to 0$
\be
\Theta(t)\sim (4\pi t)^{-n/2}e^{-t\lambda}
\sum\limits_{k=0}^\infty{(-t)^k\over k!}B_k(\lambda)
=\sum\limits_{k=0}^\infty{(-t)^k\over k!}A_k\,,
\ee
where
\be
A_k=\int\limits_M d\vol\;\tr_V\,a^{\rm diag}_k.
\ee

This is the famous Minakshisundaram-Pleijel asymptotic expansion.
The physicists call it the Schwinger-De Witt expansion \cite{barvinsky85}.
Its coefficients $A_k$ are also called sometimes
Hadamard-Minakshisundaram-De Witt-Seeley (HMDS) coefficients.
This expansion is of great importance in differential geometry, spectral
geometry,
quantum field theory and other areas of mathematical physics, such as
theory of Huygens' principle, heat kernel proofs of the index theorems,
Korteveg-De Vries hierarchy, Brownian motion etc.

One should stress, however, that this series does not converge, in general.
In that sense our ansatz (\ref{46-7/98}) or (\ref{300}) in form of a Mellin
transform of an entire
function is much better since it is exact and gives an explicit formula for the
rest term.

\subsection{Zeta Function and Determinant}

Let us apply our ansatz for computation of the complex power of the
operator $(L-\lambda)$ (with $\lambda<\lambda_0$ so that the operator
$(L-\lambda)$ is positive) defined
by
\be
G_s(\lambda)=(L-\lambda)^{-s}
={1\over\Gamma(s)}
\int\limits_0^\infty dt\, t^{s-1}\,e^{t\lambda}\,U(t).
\ee
Using our ansatz for the heat kernel one can obtain
\cite{avramidi91}
\be
G_s(\lambda)
=(4\pi)^{-n/2}\Delta^{1/2}{\cal P}{1\over 2\pi
i}\int\limits_{c-i\infty}^{c+i\infty}
dq\,{\Gamma(-q)\Gamma(-q-s+n/2)\over\Gamma(s)}
\left(\sigma\over 2\right)^{q+s-n/2}b_q(\lambda)
\ee
where $c<-{\rm Re}\,p+n/2$.

Outside the diagonal, i.e. for $\sigma\ne 0$, this integral converges for any
$s$ and defines an entire function of $s$. The integrand in this formula is a
meromorphic function of $q$ with some simple and maybe some double poles. If
we move the contour of integration to the right, we get contributions from the
simple poles in form of powers of $\sigma$ and a logarithmic part due to the
double poles (if any). This gives the complete structure of diagonal
singularities of $G_s(x,x')$.
Thus the function $b_q(\lambda)$ turns out to be very useful to study the
diagonal singularities.

Now, let us consider the diagonal limit of $G_s$. By taking the limit
$\sigma\to 0$ we
obtain a very simple formula in terms of the function $b_q$
\be
G_s^{\rm diag}(\lambda)
=(4\pi)^{-n/2}{\Gamma(s-n/2)\over\Gamma(s)}b^{\rm diag}_{n/2-s}(\lambda).
\ee
This gives automatically the zeta-function 
\be
\zeta(s,\lambda)=(4\pi)^{-n/2}
{\Gamma(s-n/2)\over\Gamma(s)}B_{n/2-s}(\lambda).
\ee

Herefrom we see that both $G_s^{\rm diag}(\lambda)$
 and $\zeta(s,\lambda)$ are meromorphic functions
of $s$
with simple poles at the points $s=[n/2]+1/2-k$, $(k=0,1,2,\dots)$ and
$s=1,2,\dots, [n/2]$.
In particular, the zeta-function is analytic at the origin. 
Its value at the
origin is given
by
\be
\zeta(0,\lambda)=\left\{
\begin{array}{ll}
0 & \mbox{ for odd } n\,,\\
(4\pi)^{-n/2}{\displaystyle {(-1)^{n/2}\over\Gamma(n/2+1)}}
B_{n/2}(\lambda) & \mbox{ for even } n\,.
\end{array}
\right.
\ee
This gives the regularized number of all modes of the operator $L$.

Moreover, the derivative of the zeta-function at the origin is also well
defined.
As we already mentioned above it determines the regularized determinant of the
operator $(L-\lambda)$
\be
\log\Det (L-\lambda)
=-(4\pi)^{-n/2}{\pi(-1)^{(n+1)/2}\over\Gamma(n/2+1)}B_{n/2}(\lambda)
\ee
for odd $n$, and
\be
\log\Det (L-\lambda)=(4\pi)^{-n/2}{(-1)^{n/2}\over\Gamma(n/2+1)}
\left\{B'_{n/2}(\lambda)-[\Psi(n/2+1)+{\bf C}]B_{n/2}(\lambda)
\right\}
\ee
for even $n$.
Here $\Psi(z)=(d/dz)\log\Gamma(z)$ is the psi-function,
${\bf C}=-\Psi(1)=0.577\dots$ is
the Euler constant, and
\be
B'_{n/2}(\lambda)={\partial\over \partial q}
B_q(\lambda)\Bigg|_{q=n/2}.
\ee

\section{Green Function}
\setcounter{equation}0

In this section we closely follow our paper \cite{avramidi98}.
Let $\lambda$ be a sufficiently large 
negative parameter, such that
$\lambda<\lambda_0$ 
and, therefore, $(L-\lambda)$ be a positive
operator. The Green function of the operator $(L-\lambda)$ reads
\be
G(\lambda|x,x')=
\sum\limits_{n=1}^\infty
\frac{1}{\lambda_n-\lambda}\varphi_n(x)\otimes \varphi^*_n(x')\,.
\ee
It is not difficult to see that 
the Green function can be represented as the Laplace transform of the
heat kernel
\be
G(\lambda)=\int\limits_0^\infty dt\;
e^{t\lambda}U(t)\,.
\ee

Using our ansatz for the heat kernel we obtain
\be
G(\lambda)
=(4\pi)^{-n/2}\Delta^{1/2}{\cal P}{1\over 2\pi
i}\int\limits_{c-i\infty}^{c+i\infty}
dq\;\Gamma(-q)\Gamma(-q-1+n/2)
\left(\sigma\over 2\right)^{q+1-n/2}b_q(\lambda)
\label{597a}
\ee
where $c<n/2-1$.

This ansatz is especially useful for studying the singularities of the Green
function, or more
general, for constructing the Green function as a power series in $\sigma$.
The integrand in (\ref{597a}) is a meromorphic function with poles
at the points $q=k$ and $q=k-1+n/2$, where
$(k=0,1,2,\dots)$.
Here one has to distinguish between odd and even dimensions.
In odd dimensions, the poles are 
at the points $q=k$ and $q=k+[n/2]-1/2$ and are simple, whereas in even
dimension there are simple poles at $q=0,1,2,\dots,n/2-2$ and 
double
poles at the points $q=k+n/2-1$.

Moving the contour of integration in (\ref{597a}) to the right one can obtain 
an expansion of the Green function
in powers of $\sigma$ (Hadamard series).
Generally, we obtain
\be
G(\lambda)=G^{\rm sing}(\lambda)
+G^{\rm non-anal}(\lambda)
+G^{\rm reg}(\lambda)\,.
\ee
Here $G^{\rm sing}(\lambda)$ is the singular part which is polynomial
in the inverse powers of $\sqrt\sigma$
\be
G^{\rm sing}(\lambda)=
(4\pi)^{-n/2}\Delta^{1/2}{\cal P}
\sum_{k=0}^{[(n+1)/2]-2}{(-1)^k\over
k!}\Gamma(n/2-k-1)\left({2\over\sigma}\right)^{n/2-k-1}
b_k(\lambda),
\ee
Let us fix an integer $N$ such that $N>(n-1)/2$.

For the rest we get in {\it odd} dimensions
\bea
&&G^{\rm non-anal}(\lambda)
+G^{\rm reg}(\lambda)
\nonumber\\
&&=
(-1)^{(n-1)/2}(4\pi)^{-{n/2}}\Delta^{1/2}{\cal P}
\sum_{k=0}^{N-(n+1)/2}
{\pi\over \Gamma\left(k+{n+1\over 2}\right)
\Gamma\left(k+{3\over 2}\right)}
\left({\sigma\over 2}\right)^{k+1/2}
b_{k+{n-1\over 2}}(\lambda)
\nonumber\\
&&
+(-1)^{(n+1)/2}(4\pi)^{-n/2}\Delta^{1/2}{\cal P}
\sum_{k=0}^{N-(n+1)/2}{\pi\over k!
\Gamma(k+n/2)}\left({\sigma\over 2}\right)^{k}
b_{k-1+n/2}(\lambda)
\nonumber\\[11pt]
&&
+(4\pi)^{-n/2}\Delta^{1/2}{\cal P}
{1\over 2\pi i}
\int\limits_{c_N-i\infty}^{c_N+i\infty}dq\,\left({\sigma\over
2}\right)^{q+1-n/2}
\Gamma(-q)\Gamma(-q-1+n/2)b_q(\lambda),
\nonumber\\
\label{597b}
\eea
where $N-1<c_N<N-1/2$.
Thus, by putting $N\to\infty$ we recover the Hadamard 
power series in $\sigma$ for {\it odd} dimension $n$
\be
G^{\rm non-anal}(\lambda)\sim
(-1)^{(n-1)/2}(4\pi)^{-{n/2}}\Delta^{1/2}{\cal P}
\sum_{k=0}^{\infty}
{\pi\over \Gamma\left(k+{n+1\over 2}\right)
\Gamma\left(k+{3\over 2}\right)}
\left({\sigma\over 2}\right)^{k+1/2}
b_{k+{n-1\over 2}}(\lambda)
\ee
\be
G^{\rm reg}(\lambda)\sim
(-1)^{(n+1)/2}(4\pi)^{-n/2}\Delta^{1/2}{\cal P}
\sum_{k=0}^{\infty}{\pi\over k!
\Gamma(k+n/2)}\left({\sigma\over 2}\right)^{k}
b_{k-1+n/2}(\lambda).
\ee

In {\it even} dimensions, the point is more subtle due to the presence of
double poles.
Moving the contour in (\ref{597a}) to the right and calculating the
contribution of the 
residues at the simple and double poles we obtain
\bea
&&G^{\rm non-anal}(\lambda)
+G^{\rm reg}(\lambda)\nonumber\\
&&=
(-1)^{n/2-1}(4\pi)^{-n/2}\Delta^{1/2}{\cal P}
\log\left(\mu^2\sigma\over 2\right)
\sum_{k=0}^{N-1}{1\over k!
\Gamma(k+n/2)}\left({\sigma\over 2}\right)^{k}
b_{k-1+n/2}(\lambda)
\nonumber\\[11pt]
&&
+(-1)^{n/2-1}(4\pi)^{-n/2}\Delta^{1/2}{\cal P}\sum_{k=0}^{N-1}
{1\over k!\Gamma(k+n/2)}\left({\sigma\over 2}\right)^{k}
\nonumber\\[11pt]
&&\times\Biggl\{
b'_{k-1+n/2}(\lambda)
-\left[\log\mu^2+\Psi(k+1)+\Psi(k+n/2)\right]
b_{k-1+{n\over 2}}(\lambda)\Biggr\}
\nonumber\\[11pt]
&&
+(4\pi)^{-n/2}\Delta^{1/2}{\cal P}
{1\over 2\pi i}\int\limits_{c_N-i\infty}^{c_N+i\infty}dq\,
\left({\sigma\over
2}\right)^{q+1-n/2}
\Gamma(-q)\Gamma(-q-1+n/2)b_{q}(\lambda),
\nonumber\\
\label{597}
\eea
where $\mu$ is an arbitrary mass parameter introduced to preserve dimensions,
$N-1<c_N<N$ and $\Psi(z)=(d/dz)\log\,\Gamma(z)$.
If we let $N\to\infty$ we obtain the Hadamard expansion of the Green function
for {\it even} dimension $n\ge 2$
\be
G^{\rm non-anal}(\lambda)\sim
(-1)^{n/2-1}(4\pi)^{-n/2}\Delta^{1/2}{\cal P}
\log\left(\mu^2\sigma\over 2\right)
\sum_{k=0}^{\infty}
{1\over k!\Gamma(k+n/2)}\left({\sigma\over 2}\right)^{k}
b_{k-1+n/2}(\lambda)
\label{597f}
\ee
\bea
G^{\rm reg}(\lambda)&\sim&
(-1)^{n/2-1}(4\pi)^{-n/2}\Delta^{1/2}{\cal P}
\sum_{k=0}^{\infty}
{1\over k!\Gamma(k+n/2)}\left({\sigma\over 2}\right)^{k}
\\
&&\times
\Biggl\{b'_{k-1+n/2}(\lambda)
-\left[\log\mu^2+\Psi(k+1)+\Psi(k+n/2)\right]
b_{k-1+{n/2}}(\lambda)\Biggr\}
\nonumber
\eea

Notice that the singular part (which is a polynomial in inverse powers of
$\sqrt\sigma$) and the {\it non-analytical} parts (proportional to
$\sqrt\sigma$ and $\log\sigma$) are expressed in terms of
the the values of the function $b_q(\lambda)$ at the
integer points $q$, 
which are uniquely locally computable from the recursion
relation,
whereas the regular
analytical part contains the values of the function $b_q(\lambda)$ at
half-integer positive points $q$ and the derivatives of the function
$b_{q}(\lambda)$ with respect to $q$ at integer positive points $q$, which are
{\it not} expressible in terms of the local information. These objects are
global and cannot be expressed further in terms of the local
heat kernel coefficients. However, they can be computed from the eqs.
(\ref{400}) and (\ref{12b}) in terms of the value of the
function $b(q)$ at some fixed point $q_0$ (see \cite{avramidi91}).

The regular part of the Green function has a well defined diagonal value 
and the functional trace. It reads in odd dimensions $n$:
\bea
\Tr\,G^{\rm reg}(\lambda)
&=&(-1)^{(n+1)/2}(4\pi)^{-n/2}{\pi\over \Gamma(n/2)}
B_{n/2-1}(\lambda)
\nonumber\\
&\stackrel{\lambda\to-\infty}{\sim} &(-1)^{(n+1)/2}(4\pi)^{-n/2}\pi
\sum_{k=0}^\infty{(-\lambda)^{n/2-1-k}\over k!\Gamma(n/2-k)}A_k
\label{25}
\eea
and in even dimensions $n$
\bea
&&\Tr\,G^{\rm reg}(\lambda)=
(-1)^{n/2-1}{(4\pi)^{-n/2}\over \Gamma(n/2)}\Biggl\{
B'_{n/2-1}(\lambda)
-\left[\log\,\mu^2+\Psi(n/2)-{\CC }\right]
B_{n/2-1}(\lambda)\Biggr\}
\nonumber\\
&&\qquad
\stackrel{\lambda\to-\infty}{\sim}
(-1)^{n/2-1}(4\pi)^{-n/2}
\Biggl\{
\sum_{k=0}^{n/2-1}{(-\lambda)^{n/2-1-k}\over k!\Gamma(n/2-k)}
\left[\CC-\Psi(n/2-k)+\log\,\left({-\lambda\over \mu^2}\right)\right]A_k
\nonumber\\
&&\qquad
+\sum_{k=n/2}^\infty{(-1)^{k-n/2}\over k!}
(-\lambda)^{n/2-1-k}\Gamma(k+1-n/2)
A_k\Biggr\}\,.
\label{33}
\eea
This trace determines the regularized vacuum expectation values like
$\left<\varphi^2\right>$ in quantum field theory.

Thus, we see that
\begin{itemize}
\item[i)]
all the singularities of the Green function and the non-analytical parts
thereof
(proportional to $\sqrt\sigma$ in odd dimensions and to $\log\sigma$ in even
dimensions)
are determined by the values of the function $b_q(\lambda)$ at
integer points $q$, which are determined, in turn, by the heat
kernel coefficients $a_k$;
\item[ii)]
there are no power singularities, i.e. $G^{\rm sing}(\lambda)=0$,
in lower dimensions $n=1,2$;
\item[iii)]
there is no logarithmic singularity (more generally, no logarithmic part at
all) in odd dimensions;
\item[iv)]
the regular part depends on the values of the function $b_q(\lambda)$
at half-integer
points $q$ and its derivative $b'_q(\lambda)$
at integer points $q$ and is a global object that
cannot be reduced
to purely local information like the heat kernel coefficients $a_k$.
\end{itemize}

The logarithmic part of the Green function
is very important.
On the one hand it determines, as usual, 
the renormalization properties of the
regular part of the Green function, i.e. the derivative
$\mu(\partial/\partial\mu)G^{\rm reg}(\lambda)$.
In particular,
\be
\mu{\partial\over\partial\mu}\Tr G^{\rm reg}(\lambda)
=\left\{
\begin{array}{ll}
0 & {\rm for\ odd\  } n\\
{\displaystyle
{(4\pi)^{-n/2}\over\Gamma(n/2)}B_{n/2-1}(\lambda)} & 
{\rm for\ even\ } n.
\end{array}
\right.
\ee

On the other hand, it is of crucial importance
in studying the Huygens principle. 
Namely, the absence of the logarithmic
part of the Green function is a necessary and sufficient
condition for the validity of the Huygens principle for hyperbolic operators.
The heat kernel coefficients coefficients and, therefore, the logarithmic
part of the Green
function are defined
for the hyperbolic operators just by analytic continuation from the elliptic
case. Thus,
the condition of the validity of Huygens principle reads
\be
\sum_{k=0}^{\infty}
{\Gamma(n/2)\over k!\Gamma(k+n/2)}\left({\sigma\over 2}\right)^{k}
b_{k-1+n/2}(\lambda)=0,
\label{597c}
\ee
or, by using (\ref{43-7/98}),
\be
\sum_{k=0}^{\infty}\sum_{j=0}^{k-1+n/2}
{\Gamma(n/2)\over k!j!\Gamma(k-j+n/2)}
\left({\sigma\over 2}\right)^{k}
(-\lambda)^{k-j}a_{j}=0\,.
\label{597d}
\ee
By expanding this equation in covariant Taylor series using the methods of 
\cite{avramidi91} one can obtain an infinite set of local conditions for
validity
of the Huygens
principle, see \cite{avramidi98}.
In particular,
\bea
&&[b_{n/2-1}(\lambda)]^{\rm diag}=0,\\
&&[\nabla_\mu b_{n/2-1}(\lambda)]^{\rm diag}=0,\\
&&[\nabla_{(\mu}\nabla_{\nu)}b_{n/2-1}(\lambda)]^{\rm diag}
+{1\over 2n}g_{\mu\nu}[b_{n/2}(\lambda)]^{\rm diag}=0\,.
\eea

%

\section{Heat Kernel Coefficients}
\setcounter{equation}0

As we have shown above the calculation of the effective action and the
Green function reduces to the calculation of the heat kernel. An important
part of that calculation is the calculation of the coefficients of
the asymptotic expansion of the heat kernel.
They are determined by a
recursion
system which is obtained simply by restricting the complex variable $q$ in the
eq.
(\ref{400-7/98})
to positive integer values $q=1,2,\dots$.

\subsection{Non-recursive Solution of the Recursion Relations}

This problem was solved in \cite{avramidi90b,avramidi91,avramidi00} where a
systematic
technique
for
calculation of $a_k$ was developed.
The formal solution of this recursion system is
\be
a_k=\left(1+{1\over k}D\right)^{-1}
\tilde L\left(1+{1\over k-1}D\right)^{-1}\tilde L\cdots
\left(1+{1\over 1}D\right)^{-1}\tilde L\cdot I.
\ee
Now, the problem is to give a precise practical meaning to this formal
operator solution.
To do this one has, first of all, to define the inverse operator
$(1+D/k)^{-1}$. This can be
done by constructing the complete set of eigenvectors of the operator $D$.
However, first we introduce some auxiliary notions from the theory of symmetric
tensors.

Let $S^n_m$ be the bundle of symmetric  tensors  of type
$(m,n)$. First of all, we define the exterior symmetric tensor product
\be
\vee: \ S^n_m\times S^i_j\to S^{n+i}_{m+j}
\ee
of symmetric tensors by  
\be
(A\vee B)_{\alpha_1\dots \alpha_{m+j}}^{\beta_1\dots \beta_{n+i}}
=A^{(\beta_1\dots \beta_n}_{(\alpha_1\dots \alpha_m}
B^{\beta_{n+1}\dots \beta_{n+i})}_{\alpha_{m+1}\dots \alpha_{m+j})}\,.
\label{vee-5/01}
\ee
Next, we define the inner product 
\be
\star:\ S^n_m\times S^i_n\to S^i_{m}
\ee
by
\be
(A\star B)_{\alpha_1\dots \alpha_{m}}^{\beta_1\dots \beta_{i}}
=A^{\gamma_1\dots \gamma_n}_{\alpha_1\dots \alpha_m}
B^{\beta_{1}\dots \beta_{i}}_{\gamma_1\dots \gamma_n}\,.
\ee
Finally, let
$\II_{(n)}$ be the identity endomorphism on the space 
of symmetric tensors of type $(n,0)$; it is a section of the bundle
$S^n_n$, that is,
\be
\II_{(n)}{}^{\mu_1\dots\mu_n}_{\nu_1\dots\nu_n}
=\delta^{(\mu_1}_{(\nu_1}\cdots\delta^{\mu_n)}_{\nu_n)}\,.
\ee

We also define the exterior symmetric covariant derivative 
\be
\nabla^S:\ S^m_n\to S^m_{n+1}
\ee 
by
\be
(\nabla^S A)_{\alpha_1\dots \alpha_{n+1}}^{\beta_1\dots \beta_m}
=\nabla_{(\alpha_1}A^{\beta_1\dots \beta_m}_{\alpha_2\dots \alpha_{n+1})}\,. 
\ee
These definitions are naturally extended to $\End({\cal V})$-valued symmetric
tensors, i.e. to the sections of the bundle $S^m_n\otimes\End({\cal V})$.

\subsection{Covariant Taylor Basis}

Let us consider the space of smooth two-point functions in a small
neighborhood of the diagonal $x=x'$; we will denote such
functions by $\left.|f\right>$.
Let us define a special set of such functions
$\{\left.|n\right>\}_{n=0}^\infty$, labeled by
a non-negative integer $n$, by 
\bea
|0\rangle &=& 1,
\\
|n\rangle &=& {1\over n!}y^{a_1}\cdots y^{a_n}\,,
\eea
where $y^a$ are the geometric parameters (normal coordinates)
defined by (\ref{ncxxz}).
These functions are scalars at the point $x$ and symmetric tensors of type
$(0,n)$ at the point $x'$. 
It is easy to show that these functions satisfy the equation
\be
D|n\rangle =n|n\rangle\,,
\ee
and, hence, are the eigenfunctions of the operator $D$ with positive integer
eigenvalues. 

Let $\langle n|$ denote the dual linear functionals defined by   
\be
\langle n|f\rangle=(\nabla^S)^nf\Big|_{x=x'}\,,
\ee
so that
\be
\langle n|m\rangle = \delta_{mn} \II_{(n)}\,,
\ee
Using this notation the covariant Taylor series for an analytic function
$|f\rangle$ can be written in the form 
\be
|f\rangle=\sum_{n=0}^\infty |n\rangle\star \langle n|f\rangle\,,
\ee

For smooth functions the Taylor series is only an asymptotic series, which
does not necessarily converge. For analytic functions, however, the Taylor
series converges in a sufficiently small neighborhood of the fixed point $x'$.
Therefore, the functions  $|n\rangle$ form a complete orthonormal basis
in the subspace of analytic functions. This is a reflection of the fact
that an analytic function that is orthogonal to all functions $|n\rangle$,
that is, whose all symmetrized derivatives vanish at the point $x'$, is, in
fact, identically equal to zero in this neighborhood.
Note, however, that the space of functions we are talking about is not a
Hilbert space since there are many analytic functions $|f\rangle$ such that
the norm $\left<f|f\right>$ defined above diverges. If we restrict ourselves 
to polynomials of some order, then this problem does not appear, and, hence,
the space of polynomials is a Hilbert space with the inner product defined
above.

\subsection{Matrix Algorithm}

The complete set of eigenfunctions $|n\rangle$ can be employed to present the
action of the operator $\tilde L$ on a function $|f\rangle$ in
the form 
\be
\tilde L|f\rangle=\sum\limits_{m,n\ge
0}|m\rangle\star\langle m|\tilde L|n\rangle \star\langle n|f\rangle, 
\label{700}
\ee
where $\langle m|\tilde L|n\rangle $ are the `matrix elements' of the operator
$\tilde L$
that are just $\End({\cal V})$-valued symmetric tensors, i.e. sections of the
vector
bundle $S^n_m\otimes \End({\cal V})$.
When acting on an analytic function this series is nothing but the Taylor
series of the result and converges in a sufficiently small neighborhood
of the point $x'$; for a smooth functions it gives an asymptotic epxansion.

Now it should be clear that the inverse of the operator
$\left(1+\frac{1}{k}D\right)^{-1}$ in
can be defined by
\be
\left(1+\frac{1}{k}D\right)^{-1}
|f\rangle=\sum_{n=0}^\infty{k\over k+n}|n\rangle \star\langle
n|f\rangle. 
\ee
Using such representations for the operators 
$\left(1+\frac{1}{k}D\right)^{-1}$
and 
$\tilde L$  we obtain a
covariant Taylor series for the coefficients $a_k$ 
\be
a_k=\sum_{n=0}^\infty|n\rangle\star\langle n|a_k\rangle
\ee
where
\bea
\langle n|a_k\rangle
&=&
\sum_{n_1,\dots,n_{k-1}\ge 0}
\frac{k}{k+n}\cdot\frac{k-1}{k-1+n_{k-1}}\cdots
\frac{2}{2+n_2}\cdot \frac{1}{1+n_1}
\nonumber\\[10pt]
&&
\times\langle n|\tilde L|n_{k-1}\rangle 
\star\langle n_{k-1}|\tilde L|n_{k-2}\rangle
\star\cdots\star \langle n_1|\tilde L|0\rangle\,,
\label{800}
\eea
where the summation is over all non-negative integers
$n_1,\dots,n_{k-1}$. It is not difficult to show that for a differential
operator of second order the matrix elements $\langle m|\tilde L|n\rangle$
do not vanish only for $n\le m+2$. Therefore, the summation over $n_i$ here is
limited from above by
\be
0\le n_1\,,\qquad
n_i\le n_{i+1}+2\,,
\qquad (i=1,2,\dots,k-1)\,,
\ee
where $n_k\equiv n$. Thus, the sum (\ref{800}) contains only a finite number
of terms.

Thus, we have reduced the problem of computation of the heat
kernel
coefficients $a_k$
to
the evaluation of the matrix elements $\langle m|\tilde L|n\rangle$ of the
operator $\tilde L$. For a differential operator $\tilde L$ of second
order, the matrix elements $\langle m|\tilde L|n\rangle $ vanish for $n> m+2$.
Therefore, the summation over $n_i$ in (\ref{800}) is limited from above:
$n_1\ge 0$, and $n_i\le n_{i+1}+2$, for  $i=1,2,\dots,k-1$, and, hence,
the sum (\ref{800}) always contains only a finite number of terms.

The matrix elements $\left<n|\tilde L|m\right>$ of a Laplace type operator have
been
computed in
our papers
\cite{avramidi91,avramidi90a}. They have the following general form
\bea
\left<m|L|m+2\right>&=&-g^{*}\vee \II_{(m)}\,,
\\[11pt]
\left<m|L|m+1\right>&=&0\,,
\\[11pt]
\left<m|L|n\right>&=&
{m\choose n}\II_{(n)}\vee Z_{(m-n)}
+{m\choose n-1}\II_{(n-1)}\vee Y_{(m-n+1)}
+{m\choose n-2}\II_{(n-2)}\vee X_{(m-n+2)}\,,
\nonumber\\
&&
\eea
where $g^{*}$ is the metric on the cotangent bundle,
$Z_{(n)}$ is a section of the vector bundle $S_n\otimes \End({\cal V})$,
$Y_{(n)}$ is a section of the vector bundle $S^1_n\otimes \End({\cal V})$ and
$X_{(n)}$ is a section of the vector bundle $S^2_n$
(a symmetric tensor of type $(2,n)$).
Here it is also meant that the binomial coefficient ${n\choose k}$ is equal to
zero if $k<0$ or $n<k$.

We will not present here explicit formulas, (they have
been computed explicitly for arbitrary $m$, $n$ in  
\cite{avramidi91,avramidi00}), but
note that all these quantities are expressed polynomially in terms of three
sorts of geometric data: 
\begin{itemize}
\item[i)]
 symmetric tensors of type $(2,n)$, i.e. sections
of the bundle $S^2_n$ obtained by symmetric derivatives
\be
K_{(n)}=(\nabla^S)^{n-2}{\rm Riem}
\ee
of the symmetrized Riemann tensor ${\rm
Riem}$ taken as a section of the bundle $S^2_2$, 
\item[ii)]
sections  
\be
{\cal R}_{(n)}=(\nabla^S)^{n-1}{\cal R}
\ee
of the vector bundle $S^1_n\otimes \End({\cal V})$
obtained by symmetrized derivatives of the curvature ${\cal R}$ of the
connection $\nabla^{\cal V}$ taken as a section of the bundle 
$S^1_1\otimes\End({\cal V})$,
\item[iii)]
$\End({\cal V})$-valued symmetric forms, i.e. sections of the vector bundle
$S^0_n\otimes\End({\cal V})$, constructed from the symmetrized
covariant derivatives 
\be
Q_{(n)}=(\nabla^S)^n Q
\ee
of the endomorphism $Q$. 
\end{itemize}


{}From
dimensional arguments it is obvious that the matrix elements
\hbox{$\left<n|L|n\right>$} are expressed in terms of the Riemann
curvature tensor,
${\rm Riem}$, the bundle curvature, ${\cal R}$, and the endomorphism $Q$;
the matrix elements $\left<n+1|L|n\right>$ ---
in terms of the quantities $\nabla {\rm Riem}$, $\nabla{\cal R}$ and $\nabla
Q$;
the elements $\left<n+2|L|n\right>$ ---
in terms of the quantities of the form
$\nabla\nabla {\rm Riem}$, ${\rm Riem}\cdot {\rm Riem}$,
etc.

\subsection{Diagramatic Technique}

In the computation of the heat kernel 
coefficients by means of the matrix
algorithm
a ``diagrammatic'' technique, i.e., a graphic method for
enumerating the different terms of the sum (\ref{800}), turns out to be very
convenient and pictorial \cite{avramidi91,avramidi00}.

The matrix elements $\left<m|L|n\right>$ are presented by some blocks with $m$
lines coming in from the left and $n$ lines going out to the right (Fig. 1),
\begin{center}
\unitlength1mm
\begin{picture}(41,10)
\put(-2.0,-1.0){$m\,\Biggl\{$}
\put(5.0,4.0){\line(1,0){12}}
\put(7.0,0.0){$\vdots$}
\put(5.0,-1.5){\line(1,0){10}}
\put(5.0,-4.0){\line(1,0){12}}
\put(20.0,0){\circle{10}}
\put(23.0,4.0){\line(1,0){12}}
\put(32.0,0.0){$\vdots$}
\put(25.0,-1.5){\line(1,0){10}}
\put(23.0,-4.0){\line(1,0){12}}
\put(36.0,-1.0){$\Biggl\}\,n$}
\end{picture}
\end{center}
\vglue4mm
\begin{center}
Fig. 1
\end{center}
and the product of the matrix elements
$\left<m|L|k\right>\star\left<k|L|n\right>$ --- by two blocks
connected by $k$ intermediate lines (Fig. 2),

\begin{center}
\unitlength1mm
\begin{picture}(61,10)
\put(-2.0,-1.0){$m\,\Biggl\{$}
\put(5.0,4.0){\line(1,0){12}}
\put(7.0,0.0){$\vdots$}
\put(5.0,-1.5){\line(1,0){10}}
\put(5.0,-4.0){\line(1,0){12}}
\put(20.0,0){\circle{10}}
\put(23.0,4.0){\line(1,0){14}}
\put(25.0,-1.5){\line(1,0){10}}
\put(23.0,-4.0){\line(1,0){14}}
\put(26.0,-1.0){$k\,\Biggl\{$}
\put(32.0,0.0){$\vdots$}
\put(40.0,0){\circle{10}}
\put(43.0,4.0){\line(1,0){12}}
\put(52.0,0.0){$\vdots$}
\put(45.0,-1.5){\line(1,0){10}}
\put(43.0,-4.0){\line(1,0){12}}
\put(56.0,-1.0){$\Biggl\}\,n$}
\end{picture}
\end{center}
\vglue4mm
\begin{center}
Fig. 2
\end{center}
that represents the contractions of the corresponding tensor indices (the inner
product).

To obtain the coefficient $\left<n|a_k\right>$ one should draw, first, all
possible
diagrams which have $n$ lines incoming from the left and which are
constructed from $k$ blocks connected in all possible ways by any number of
intermediate lines.
When doing this, one should keep in mind that the number of
the lines, going out of any block, cannot be greater than the number of the
lines,
coming in, by more than two and by exactly one.
Then one should
sum up all diagrams with the weight determined for each diagram by the number
of intermediate lines from the analytical formula (\ref{800}).
Drawing of such diagrams is of no difficulties.
This helps to keep under control the whole variety of different terms.
Therefore, the main problem is reduced to the computation of some standard
blocks,
which can be computed once and for all.

For example, the diagrams for the diagonal values of the HMDS-coefficients
$a_k^{\rm diag}=$ \mbox{$\left<0|a_k\right>$} have the form,

\unitlength1mm
\newsavebox{\bone}
\savebox{\bone}(22,1.5)[l]
{\multiput(4,2)(7,0){1}{\circle{4}}
\put(19.35,3.5){}}
\be
a_1^{\rm diag}=\usebox{\bone}
\ee

\unitlength1mm
\newsavebox{\btwoa}
\savebox{\btwoa}(22,1.5)[l]
{\multiput(4,2)(7,0){2}{\circle{4}}
\put(19.35,3.5){}}
\newsavebox{\btwob}
\savebox{\btwob}(22,1.5)[l]
{\multiput(4,2)(7,0){2}{\circle{4}}
\put(5.3,3.5){\line(1,0){4.37}}
\put(5.3,0.5){\line(1,0){4.37}}}
\be
a^{\rm diag}_2=\usebox{\btwoa}
\!\!\!\!\!\!\!\!\!\!
+{1\over 3}\usebox{\btwob}
\ee

%
%

\unitlength1mm
\newsavebox{\fblocka}
\savebox{\fblocka}(22,1.5)[l]
{\multiput(4,2)(7,0){3}{\circle{4}}
\put(19.35,3.5){}}
\newsavebox{\fblockb}
\savebox{\fblockb}(22,1.5)[l]
{\multiput(4,2)(7,0){3}{\circle{4}}
\put(12.4,3.5){\line(1,0){4.37}}
\put(12.4,0.5){\line(1,0){4.37}}}
\newsavebox{\fblockc}
\savebox{\fblockc}(22,1.5)[l]
{\multiput(4,2)(7,0){3}{\circle{4}}
\put(5.3,3.5){\line(1,0){4.37}}
\put(5.3,0.5){\line(1,0){4.37}}}
\newsavebox{\fblockd}
\savebox{\fblockd}(22,1.5)[l]
{\multiput(4,2)(7,0){3}{\circle{4}}
\put(5.3,3.5){\line(1,0){4.37}}
\put(5.3,0.5){\line(1,0){4.37}}
\put(12.9,2){\line(1,0){3.1}}}
\newsavebox{\fblocke}
\savebox{\fblocke}(22,1.5)[l]
{\multiput(4,2)(7,0){3}{\circle{4}}
\put(5.3,3.5){\line(1,0){4.37}}
\put(5.3,0.5){\line(1,0){4.37}}
\put(12.4,3.5){\line(1,0){4.37}}
\put(12.4,0.5){\line(1,0){4.37}}}
\newsavebox{\fblockf}
\savebox{\fblockf}(22,1.5)[l]
{\multiput(4,2)(7,0){3}{\circle{4}}
\put(5.3,3.5){\line(1,0){4.37}}
\put(5.3,0.5){\line(1,0){4.37}}
\put(12.35,3.5){\line(1,0){4.37}}
\put(12.35,0.5){\line(1,0){4.37}}
\put(12.8,2.5){\line(1,0){3.2}}
\put(12.8,1.5){\line(1,0){3.2}}}
\begin{eqnarray}
a^{\rm diag}_3&=&\usebox{\fblocka}
+{1\over 3}\usebox{\fblockb}
+{2\over 4}\usebox{\fblockc}
\\[10pt]
& &
+{2\over 4}\cdot{1\over 2}\usebox{\fblockd}
+{2\over 4}\cdot{1\over 3}\usebox{\fblocke}
+{2\over 4}\cdot{1\over 5}\usebox{\fblockf}\,.
\nonumber
\end{eqnarray}

\null
\par
\medskip
As an illustration let us  compute the coefficients 
$a^{\rm diag}_1$ and $a^{\rm diag}_2$. We
have
\cite{avramidi91}

\newsavebox{\bonenew}
\savebox{\bonenew}(7,1.5)[l]
{\multiput(4,2)(7,0){1}{\circle{4}}
\put(19.35,3.5){}}

\be
\usebox{\bonenew}=\left<0|L|0\right>=Z_{(0)}
\ee

\newsavebox{\btwobnew}
\savebox{\btwobnew}(15,1.5)[l]
{\multiput(4,2)(7,0){1}{\circle{4}}
\put(5.3,3.5){\line(1,0){4.37}}
\put(5.3,0.5){\line(1,0){4.37}}}
\be
\usebox{\btwobnew}=\left<0|L|2\right>=-g^{ab}
\ee

\newsavebox{\btwobnewag}
\savebox{\btwobnewag}(15,1.5)[l]
{\multiput(11,2)(7,0){1}{\circle{4}}
\put(5.3,3.5){\line(1,0){4.37}}
\put(5.3,0.5){\line(1,0){4.37}}}

\be
\usebox{\btwobnewag}=\left<2|L|0\right>=Z_{(2)ab}
\ee

\be
\usebox{\btwob}\hspace{-10pt}
=\left<0|L|2\right>\star\left<2|L|0\right>
=-g^{ab}Z_{(2)ab}\,,
\ee
where
\be
Z_{(0)}=Q-{1\over 6}R\,\,\II_V\,,
\ee
\bea
Z_{(2)ab}&=&
\nabla_{(a}\nabla_{b)}Q
-{{1}\over {2}}{\cal R}_{c(a}{\cal R}^c{}_{b)}
+{1\over 2}\nabla_{(a}\nabla_{|c|}{\cal R}^c{}_{b)}
\\[10pt]
&&
+\II_V\,\Biggl(-{3\over 20}\nabla_a\nabla_b R
-{1\over 20}\Delta R_{ab}
+{{1}\over {15}}R_{ac}R^c{}_{b}
-{{1}\over {30}}R_{acde}R_b{}^{cde}
-{1\over 30}R_{cd}R^c{}_a{}^d{}_b{}\Biggr)\,.
\nonumber
\eea
Here, as usual, the parenthesis denote the complete symmetrization
over all indices included and
the vertical lines indicate the indices excluded from 
the symmetrization.
Hence, we immediately get
\be
a^{\rm diag}_1=Q-{1\over 6}R\,\,\II_V,
\ee
and, by taking the trace of $Z_{(2)}$ and using the identity
$\nabla_a\nabla_b{\cal R}^{ab}=0$,
we obtain the well known result \cite{avramidi91}
\be
a^{\rm diag}_2=\left(Q-{1\over 6}R\,\II_V\right)^2
-{1\over 3}\Delta Q
+{1\over 6}{\cal R}_{ab}{\cal R}^{ab}
+\II_V\,\Biggl({1\over 15}\Delta R
-{1\over 90} R_{ab}R^{ab}
+{{1}\over {90}}R_{abcd}R^{abcd}\Biggr)\,.
\ee


The technique described above is manifestly covariant and
is applicable for any Riemannian (or pseudo-Riemannian) manifold $M$
and for any vector bundle $V$.
It is also valid for local analysis on 
noncompact manifolds and manifolds with boundary 
(on a finite distance from
the boundary).
This method gives not only the diagonal values of the heat
kernel coefficients but also
the diagonal values of all their derivatives, that is, it gives 
also the off-diagonal coefficients in form of a covariant Taylor series.
Due to the use of symmetric forms and symmetric covariant
derivatives the famous `combinatorial explosion' in the complexity of the
heat kernel coefficients
is avoided.
This technique is very algorithmic and well suited to
automated computation.
The developed method is very powerful; it enabled us to compute for the
first time
the diagonal value of the fourth HMDS-coefficient $a^{\rm diag}_4$
\cite{avramidi90b,avramidi91}. It was used in 
\cite{yajima04,vandeven98} to compute
the coefficient $a^{\rm diag}_5$.
Lastly, this technique enables one not only to carry out explicit computations,
but
also to analyse the general structure of the 
heat kernel coefficients for
all orders $k$.

\subsection{General Structure of Heat Kernel Coefficients}

Now we are going to investigate the general structure of the 
heat kernel coefficients.
We will follow mainly our papers
\cite{avramidi00,avramidi99a,avramidi02}.

Our analysis will be again purely local.
Since locally one can always expand the metric,
the connection and the endomorphism $Q$ in the covariant Taylor series,
they are completely characterized by their Taylor coefficients, i.e.
the covariant derivatives of the curvatures, more precisely by the objects
$K_{(n)}$, ${\cal R}_{(n)}$ and $Q_{(n)}$ introduced above.
We introduce the following notation for all of them
\be
\Re_{(n)}=\{K_{(n+2)}, {\cal R}_{(n+1)}, Q_{(n)}\},
\qquad (n=0,1,2,\dots),
\ee
and call these objects {\it covariant jets};
$n$ will be called the order of a jet $\Re_{(n)}$.
It is worth noting that the jets are defined by {\it symmetrized} covariant
derivatives. This makes them well defined as the order of the derivatives
becomes not important. It is only the {\it number} of derivatives that plays
a role.

The low-order coefficients $A_0$ and $A_1$ have been described
above.
As far as the higher order coefficients $A_k$, $(k\ge 2)$, are concerned
they are integrals of local invariants which are polynomial in the jets.
One can classify all the terms in them according to the number of
the jets and their order.
The terms linear in the jets in higher order coefficients
$A_k$, ($k\ge 2$), are given by integrals of total derivatives,
symbolically $\int_Md\vol\,\tr_V\,\Delta^{k-1}\Re$. They are calculated
explicitly
in \cite{avramidi91,avramidi00,avramidi90a}.
Since the total derivative do not contribute to an integral
over a complete compact manifold,
it is clear that the linear terms vanish.
Thus $A_k$, $(k=2,3,\dots)$, begin with the terms quadratic in the jets.
These terms contain the jets of highest order (or the {\it leading derivatives}
of the curvatures)
and can be shown to be of the form 
$\int_Md\vol\;\tr_V\,\Re \Delta^{k-2}
\Re$.
Then it follows a class of terms cubic in the jets etc.
The last class of terms does not contain any
covariant derivatives at all but only the powers of the curvatures.
In other words, the higher order HMDS-coefficients have a general structure,
which can be presented symbolically in the form

Thus, for $k\ge 2$  one can classify the terms
in $A_{2k}$ according to the number of the jets and their order
\be
A_{k}=\sum_{j=2}^k A_{k,(j)}\,,
\ee
where $A_{k,(j)}$ is the contribution of order 
$j$ in the jets; they can be presented symbolically in the form
\bea
A_{k,(2)} &=&
\int\limits_M d\vol\;{\rm tr}_V\,\sum\Re_{(0)}\Re_{(2k-4)},
\\
A_{k,(3)} &=& \int\limits_M d\vol\;{\rm tr}_V\,
\sum_{i=0}^{2k-6}\sum \Re_{(0)}\Re_{(i)}\Re_{(2k-6-i)},
\\
&\cdots&
\nonumber\\
A_{k,(k-1)} &=& \int\limits_M d\vol\;{\rm tr}_V\,
\left[
\sum\Re_{(2)}\left(\Re_{(0)}\right)^{k-2}
+\sum\left(\Re_{(1)}\right)^2
\left(\Re_{(0)}\right)^{k-3}\right]\,,
\\
A_{k,(k)} &=& \int\limits_M d\vol\;{\rm tr}_V\,\sum
\left(\Re_{(0)}\right)^k.
\eea
More precisely, the functionals $A_{k,(j)}$ transform under the rescaling
of the jets
\be
\Re_{(k)} \mapsto \varepsilon \alpha^k \Re_{(k)} 
\ee
as follows
\be
A_{k,(j)}\mapsto \varepsilon^j\alpha^{2(k-j)} A_{k,(j)}\,.
\ee

\section{High Energy Approximation}
\setcounter{equation}0

One can show that 
all quadratic terms can be reduced to five independent invariants, viz.
\cite{avramidi90a,avramidi91,avramidi00}
\begin{eqnarray}
A_{k,(2)}&=&
{k!(k-2)!\over 2(2k-3)!}
\int\limits_M d\vol\;{\rm tr}_V\,
\Biggl\{f^{(1)}_kQ\Delta^{k-2}Q
+2f^{(2)}_k{\cal R}^{bc}\nabla_b
\Delta^{k-3}\nabla_a{\cal R}^a{}_c
\nonumber\\[10pt]
&&
+f^{(3)}_k Q\Delta^{k-2}R
+f^{(4)}_kR_{ab}\Delta ^{k-2}R^{ab}
+f^{(5)}_kR\Delta^{k-2}R
\Biggr\},
\label{4-5-3.53}
\end{eqnarray}
where $f^{(i)}_k$ are some numerical coefficients.
These numerical coefficients can be computed by the technique developed in the
previous section.
{}From the formula
(\ref{800}) we have for the diagonal coefficients $a_k^{\rm diag}$
up to cubic terms in
the jets
\begin{eqnarray}
a_k^{\rm diag}
&=&\left<0|a_k\right>
={(-1)^{k-1}\over {2k-1\choose k}}\left<0;k-1|L|0\right>
\nonumber\\[10pt]
&&
+(-1)^k\sum\limits_{i=1}^{k-1}\,\sum\limits_{n_i=0}^{2(k-i-1)}
{{2k-1\choose i}\over {2k-1\choose k}{2i+n_i-1\choose i}}
\left<0;k-i-1|L|n_i\right>\star\left<n_i;i-1|L|0\right>
\nonumber\\[11pt]
&&
+O(\Re^3),
\label{4-5-3.52}
\end{eqnarray}
where
\be
\left<n;k|L|m\right>=(\vee^k g^{*})\star\left<n|L|m\right>
\ee
and $O(\Re^3)$ denote terms of third order in the jets.

By computing the matrix elements in the second order in the jets and
integrating over $M$ one obtains \cite{avramidi90a,avramidi91}
\bea
f^{(1)}_k&=&1\,,\\
f^{(2)}_k&=&{{1}\over {2(2k-1)}}\,,\\
f^{(3)}_k&=&{{k-1}\over {2(2k-1)}}\,,\\
f^{(4)}_k&=&{{1}\over {2(4k^2-1)}}\,,\\
f^{(5)}_k&=&{{k^2-k-1}\over {4(4k^2-1)}}\,.
\label{4-5-4.7}
\eea
One should note that the same results were obtained by a completely different
method in \cite{branson90b}.


Let us consider the situation when the curvatures are small but rapidly
varying ({\it high energy approximation} in quantum field theory), i.e.
the derivatives of the curvatures are more important than the powers of them.
This corresponds to an asymptotic expansion in the
deformation parameter $\varepsilon$ as $\varepsilon\to 0$.
Then the leading derivative terms in the heat kernel are the largest ones.
Thus the heat trace has the form
\be
\Theta(t)\sim
(4\pi t)^{-n/2}\left\{A_0-tA_1+{t^2\over 2}H_2(t)\right\}+O(\Re^3),
\ee
where $H_2(t)$ is some complicated {\it nonlocal} functional
that has the following asymptotic expansion as $t\to 0$
\be
H_2(t)\sim 2\sum_{k=2}^\infty
{(-t)^{k-2}\over k!}A_{k,(2)}
\,.
\ee
Using the results for $A_{k,(2)}$ one can easily construct such a functional
$H_2$ just by a formal summation of the leading derivatives
\bea
H_2(t)&=&\int\limits_M d\vol\;\tr_V\,\Biggl\{
Q\gamma^{(1)}(-t\Delta)Q
+2{\cal R}^a{}_{c}\nabla_a{{1}\over {\Delta}}
\gamma^{(2)}(-t\Delta)\nabla_b{\cal R}^{bc}
\nonumber\\
&&-2Q\gamma^{(3)}(-t\Delta)R
+R_{ab}\gamma^{(4)}(-t\Delta)R^{ab}
+R\gamma^{(5)}(-t\Delta)R
\Biggr\},
\label{4-5-4.13}
\eea
where $\gamma^{(i)}(z)$ are entire functions defined by
\cite{avramidi90a,avramidi91}
\be
\gamma^{(i)}(z)=\sum\limits_{k=0}^\infty
{k!\over (2k+1)!}f^{(i)}_kz^k
=\int\limits_0^1 d\xi\, f^{(i)}(\xi)
\exp\left(-{{1-\xi^2}\over {4}}z\right)\,,
\label{4-5-4.12}
\ee
where
\bea
f^{(1)}(\xi)&=&1\,,\\
f^{(2)}(\xi)&=&{{1}\over {2}}\xi^2\,,\\
f^{(3)}(\xi)&=&{{1}\over {4}}(1-\xi^2)\,,\\
f^{(4)}(\xi)&=&{{1}\over {6}}\xi^4\,,\\
f^{(5)}(\xi)&=&{{1}\over {48}}(3-6\xi^2-\xi^4)\,.
\label{4-5-4.9}
\eea
Therefore, $H_2(t)$ can be regarded as generating functional for
quadratic terms
$A_{k,(2)}$ (leading derivative terms) in all coefficients $A_k$.
It also plays a very important role in investigating the nonlocal structure of
the effective
action in quantum field theory in high-energy approximation 
\cite{avramidi90a,avramidi91}.


%
\section{Low Energy Approximation}
\setcounter{equation}0

Let us consider now the opposite case, when the curvatures are strong but
slowly
varying
(low-energy approximation in quantum field theory), 
i.e. the powers of the
curvatures are more important than the
derivatives of them.
This corresponds to the asynptotic expansion in the deformation parameter
$\alpha$ as $\alpha\to 0$.
The main terms in this approximation are the terms without any covariant
derivatives
of the curvatures, i.e. the lowest order jets.
We will consider mostly the zeroth order of this approximation
which corresponds simply to covariantly constant background curvatures
\be
\na {\rm Riem} = 0,\qquad \na {\cal R}=0,
\qquad \na Q = 0.
\ee

The asymptotic expansion of the heat trace
\be
\Theta(t)\sim (4\pi t)^{-n/2}\sum_{k=0}^\infty{(-t)^k\over
k!}A_{k,(k)}.
\ee
determines then {\it all} the terms without covariant derivatives
(highest order terms in the jets), $A_{k,(k)}$, in all 
heat kernel coefficients $A_k$.
These terms do not contain any covariant derivatives and are just
polynomials in the curvatures and the endomorphism $Q$.
Thus the heat trace is a generating
functional for all heat kernel 
coefficients for a covariantly constant
background, in particular, for all symmetric spaces.
Thus the problem is to calculate  the heat trace 
for covariantly constant background.

\subsection{Algebraic Approach}

There exist a very elegant indirect way to construct the heat
kernel without solving the heat equation but using only the
commutation relations of some covariant first order differential
operators. Below we wil follow our papers
\cite{avramidi93,avramidi94a,avramidi95c,avramidi96,avramidi08a,avramidi08b}.
The main idea is to employ a generalization of the usual Fourier
transform to the case of operators; it consists in the following.
We are going to use the following representation of the heat trace
\be
\Theta(t)=\int\limits_M d\vol\;\tr_V
\left[\exp(-tL)\delta(x,x')\right]^{\rm diag}\,.
\ee 

Let us consider for a moment a trivial case, where the
curvatures vanish but the potential term does not:
\be
{\rm Riem}= 0,\qquad {\cal R}=0,\qquad \nabla Q=0.
\ee
In this case the operators of covariant derivatives obviously
commute and form together with the potential term an Abelian
Lie algebra
\be
[\na_\mu,\na_\nu]=0,\qquad [\na_\mu, Q]=0.
\ee
It is easy to show that the heat semigroup can
be presented in the form
\be
\exp(-tL)=(4\pi t)^{-n/2}\exp(-tQ)\int\limits_{\RR^n} dk\;g^{1/2}
\exp\left(-{1\over 4t}\left<k,gk\right>
+k\cdot\na\right),
\label{1300}
\ee
where $\left<k,gk\right>=k^\mu g_{\mu\nu}k^\nu$ and
$k\cdot\nabla=k^\mu
\nabla_\mu$.
Here, of course, it is assumed that the covariant derivatives also
commute with the metric
\be
[\nabla, g]=0.
\ee
Acting with this operator on the Dirac distribution and using the
obvious relation
\be
\left[
\exp(k\cdot\na)\delta(x,x')\right]^{\rm diag}=\delta(k),
\ee
one can integrate easily over $k$ to obtain the heat
trace
\be
\Theta(t)=(4\pi t)^{-n/2}\int\limits_M d\vol\;\tr_V\exp(-tQ).
\ee

Of course, on curved manifolds the covariant differential operators
$\nabla$ do not commute---their commutators are determined by
the curvatures $\Re$.
The commutators of covariant derivatives $\nabla$ with the curvatures
$\Re$ give the first
derivatives of the curvatures, i.e. the jets $\Re_{(1)}$, the
commutators of
 covariant derivatives with $\Re_{(1)}$ give the second jets
 $\Re_{(2)}$, etc.
Thus the operators $\nabla$ together with the
whole set of the jets ${\cal J}$ form an infinite dimensional
Lie algebra
${\cal G}=\{\nabla, \Re_{(i)}; (i=1,2,\dots)\}$.

Now, let us remember that the heat trace is a functional of the
jets, with the jets being defined by symmetrized covariant derivatives.
This makes the order of a jet well defined. For example, the
structures involving commutators of covariant derivatives, 
(like $[\nabla_a,\nabla_b]R^e{}_c{}^f{}_d$, 
which involve 2-jets of the Riemann tensor on the left but, 
after using the Ricci identity, only $0$-jets on the right) 
are not allowed. After symmetrizing over $abcd$ this jet vanishes.
So, if we express the final answer for the heat kernel diagonal or
for the heat kernel coefficients in terms of the symmetrized jets,
then there is a natural filtration with respect to the
order of the jets involved. In other words, one can always say, what is
the maximal order of symmetrized covariant derivative of the
curvature involved in the result. This is especially true for the heat kernel
coefficients $A_k$ since they are polynomial in the jets.

If we identify a small deformation parameter $\alpha$ 
with each derivative then a jet
of order $n$ is, actually, of order $\alpha^n$.
Thus, we get a perturbation theory in this small parameter.
Since the derivatives are naturally identified with the momentum (or energy),
the physicists call a situation when the derivatives are small 
the {\it low-energy approximation}. 
To evaluate the heat kernel in the low-energy approximation one
can take into account only a finite number of low-order jets,
i.e. the
low-order covariant derivatives of the background fields,
$\{\Re_{(i)}; (i\le N)\}$ with some fixed $N$,  and neglect all
higher order jets, i.e. the covariant derivatives of
higher orders,
i.e. put $\Re_{(i)}=0$\ for $i> N$.
Then one can show that there exist a set of covariant differential
operators
that together with the low-order jets
generate a finite-dimensional Lie algebra
${\cal G}_N=\{\nabla, \Re_{(i)};
(i=1,2,\dots,N)\}$.
One should stress here what problem one can solve this way.
We try to answer the following concrete 
question: how do the heat kernel coefficients
look if we through away all the (symmetrized)
jets of order higher than $N$?

Thus one can try to generalize
the above idea in such a way that (\ref{1300}) would be the zeroth
approximation
in the commutators of the covariant derivatives, i.e. in the
curvatures.
Roughly speaking, we would like to find a representation of the
heat semigroup
in the form
\be
\exp(-tL)=(4\pi t)^{-D/2}
\int\limits_{\RR^D} dk\,
\Phi(t,k)\exp\left(-{1\over 4t}\left<k,\Psi(t) k\right>
+T(k)\right)\,,
\ee
where $\left<k,\Psi(t) k\right>=k^A\Psi_{AB}(t)k^B$,
$T(k)=k^A T_A$, ($A=1,2,\dots,D$),
$T_A$ are some first order differential operators
and the functions $\Psi(t)$ and $\Phi(t,k)$ are expressed in
terms  of commutators of these
operators, i.e., in terms of the curvatures;
that is, these functions are analytic functions of $t$.
In general, the operators $T_A$ do not form a closed finite
dimensional algebra because at each step, by taking more
commutators, there appear more and more derivatives of the
 curvatures. It is the  low-energy reduction
${\cal G}\mapsto {\cal G}_N$, i.e. the restriction to the
low-order jets, that actually closes the algebra ${\cal G}$
of the operators $T_A$ and the background jets, i.e. makes
it finite dimensional.

Using this representation one can, as above, act with
$\exp\left[T(k)\right]$ on the Dirac distribution to get the
heat kernel. The main point of this idea is that it is
much easier to calculate the action of the exponential
of the first order operator $T(k)$ on the Dirac
distribution
than that of the exponential of the second
order operator $L$.

\subsection{Covariantly Constant Background in Flat Space}

Let us consider now the more complicated case of nontrivial
covariantly constant curvature of the connection on the vector bundle $V$
in flat space:
\be
{\rm Riem}= 0,\qquad \nabla{\cal R}=0,
\qquad \nabla Q=0.
\ee

Using the condition of covariant constancy of the curvatures
one can show that in this case the covariant derivatives form
a {\it nilpotent} Lie algebra \cite{avramidi93}
\bea
&&[\nabla_\mu,\nabla_\nu]={\cal R}_{\mu\nu}, \\
&&[\nabla_\mu,{\cal R}_{\alpha\beta}]=[\nabla_\mu,Q]=0,\\
&&[\R _{\mu\nu},\R _{\a\b}]=[\R _{\mu\n},Q]=0.
\eea

For this algebra one can prove a theorem expressing the heat
semigroup operator in terms of an average over the
corresponding Lie group
\cite{avramidi93}
\bea
\exp(-tL)&=&(4\pi t)^{-n/2}\exp(-tQ)
\left[\det{}_{TM}\left({t\R \over \sinh(t\R )}\right)\right]^{1/2}
\\
&&\times\int\limits_{\RR^n}dk\; g^{1/2}
\exp\left(-{1\over 4t}\left<k, g t\R  \coth(t\R ) k\right>
+k\cdot\nabla\right),
\eea
where $k\cdot\nabla=k^\mu\nabla_\mu$.
Here functions of the curvatures $\R $ are understood as
functions of sections of the bundle $\End(TM)\otimes\End({\cal V})$,
and the determinant
$\det_{TM}$ is taken with respect to the tangent space
indices; the fiber indices of the bundle   
${\cal V}$ being intact.

It is not difficult to show that in this case we also have
\be
\left[\exp\left(k\cdot\nabla\right)\delta(x,x')\right]^{\rm diag}
=\delta(k)\,.
\ee
Subsequently, the integral over $k$ becomes trivial and
we obtain immediately the trace of the heat kernel \cite{avramidi93}
\be
\Theta(t)=(4\pi t)^{-n/2}\int\limits_M d\vol\;\tr_V\exp(-tQ)
\left[\det{}_{TM}\left({t\R  \over \sinh(t\R )}\right)
\right]^{1/2}\,.
\ee
Expanding it in a power series in $t$ one can find {\it all}
covariantly constant terms in {\it all} heat kernel 
coefficients $A_k$.

As we have seen the contribution of the curvature
$\R _{\mu\n}$ is not as trivial as that of the potential term.
However, the algebraic approach does work in this case too.
It is a good example how one can get the heat kernel without
solving any differential equations but using only the algebraic
properties of the covariant derivatives.

\subsubsection{Quadratic Potential in Flat Space}

In fact, in flat space it is possible to do a bit more, i.e.
to calculate the contribution of the first and the second
derivatives of the potential term $Q$
\cite{avramidi95c}.
That is, we consider the case when the derivatives of the
endomorphism $Q$ vanish only starting from the {\it third}
order, i.e.
\be
{\rm Riem}= 0,\qquad \nabla{\cal R}=0,\qquad
\nabla\nabla\nabla Q=0.
\ee
Besides we assume the background to be {\it Abelian},
i.e. all the nonvanishing background quantities,
$\R _{\a\b}$, $Q$, $Q_{;\mu}\equiv\nabla_\mu Q$ and
$Q_{;\n\mu}\equiv \nabla_\mu\nabla_\nu Q$, commute with each other.
Thus we have again a nilpotent Lie algebra
\bea
&&[\nabla_\mu, \nabla_\nu]={\cal R}_{\mu\nu}\,,\\
&&[\nabla_\mu,Q]=Q_{;\m}\,,\\
&&[\nabla_\m,Q_{;\n}]=Q_{;\n\m}\,,
\label{1400}
\eea
all other commutators being zero.

Now, let us represent the endomorphism $Q$ in the form
\be
Q=Q_0-\a^{ik}N_iN_k,
\ee
where $(i=1,\dots,q; q\le n),$ $\a^{ik}$ is some constant
symmetric nondegenerate $q\times q$ matrix, $Q_0$ is a
covariantly constant endomorphism and $N_i$ are some endomorphisms with
vanishing second covariant derivative:
\be
\nabla Q_0=0, \qquad \nabla\nabla N_i=0.
\ee
Next, let us introduce the operators $X_A=(\nabla_\m, N_i)$,
$(A=1,\dots, n+q)$ and the matrix
\be
(\F _{AB})=\left(\matrix{
\R_{\m\n} & N_{i;\m}\cr
-N_{k;\n} & 0       \cr}\right),
\ee
with $N_{i;\m}\equiv \nabla_\m N_i$.

The operator $L$ can now be written in the form
\be
L=-G^{AB}X_A X_B+Q_0\,,
\ee
where
\be
(G^{AB})=\left(\matrix{g^{\m\n} & 0 \cr
			0 & \a^{ik} \cr}\right).
\ee
and the commutation relations (\ref{1400}) take a more compact form
\be
[X_A, X_B]=\F _{AB}\,,
\ee
all other commutators being zero.

This algebra is again a nilpotent Lie algebra.
Thus one can apply the previous results  in this case too to get
\cite{avramidi95c}
\bea
\exp(-tL)&=&(4\pi t)^{-{(n+q)}/2}\exp(-tQ_0)
\left[\det\left({t\F \over \sinh(t\F )}\right)\right]^{1/2}
\nonumber\\
&&
\times\int\limits_{\RR^{n+q}} d k\; G^{1/2}
\exp\left(-{1\over 4t}\left<k, G t\F  \coth(t\F )k\right>
+X(k)\right),
\eea
where $G=\det G_{AB}$ and $X(k)=k^A X_A$.

Thus we have expressed the heat semigroup operator in terms
of the operator
\hbox{$\exp\left[X(k)\right]$}. The integration over $k$ is
Gaussian except for the noncommutative part.
Splitting the integration variables $(k^A)=(q^\mu,\omega^i)$ and
using the
Campbell-Hausdorf formula we obtain \cite{avramidi95c}
\be
\left[\exp\left[X(k)\right]\delta(x,x')\right]^{\rm diag}
=\exp\left[N(\omega)\right]\delta(q),
\ee
where $N(\omega)=\omega^i N_i$.
Further, after taking off the trivial
integration over $q$
and a Gaussian integral over $\om$,
we obtain the heat trace \cite{avramidi95c}.

To describe the result let us introduce 
a matrix determined by second derivatives of the
potential term as follows
\be
P=\left(P^\m{}_{\n}\right), \qquad
P^\m{}_{\n}={1\over 2} \nabla^\mu\nabla_\nu Q,
\ee
and
the matrices $C(t)=(C^\m{}_{\n}(t))$, $K(t)=(K^\m{}_{\n}(t))$
$S(t)=(S^\m{}_{\n}(t))$ and $E(t)=(E^\m{}_{\n}(t))$ 
by
\be
C(t)=\oint\limits_C{dz\over 2\pi i}\,F(z)\,
t\coth\left(\frac{t}{z}\right)\,,
\ee
\be
K(t)=\oint\limits_C{dz\over 2\pi i}\,F(z)
{t\over z^{2}}\sinh\left(\frac{t}{z}\right)\,,
\ee
\be
S(t)=\oint\limits_C{dz\over 2\pi i}\,F(z){t\over z}\,
\sinh\left(\frac{t}{z}\right)\,,
\ee
\be
E(t)=\oint\limits_C{dz\over 2\pi i}F(z)\,t\,
\sinh\left(\frac{t}{z}\right)\,,
\ee
where 
\be
F(z)=(1-z\R -z^2 P)^{-1}
\ee
and the integral is taken along a sufficiently small closed contour $C$
that encircles the origin counter-clockwise, so that $F(z)$
is analytic inside this contour.

Then the heat trace has the form
\be
\Theta(t)=(4\pi t)^{-n/2}\int\limits_Md\vol\;\tr_V
\left[\Phi(t)\right]^{-1/2} \exp\left[-tQ
+{1\over  4} t^3 \left<\nabla Q, \Psi(t)\nabla Q\right>\right],
\label{1500}
\ee
where
$
\left<\nabla Q, \Psi(t)\nabla Q\right>
=\nabla_\m Q\Psi^\m{}_{\n}(t)\nabla^\nu Q,
$
\bea
\Phi(t)&=&
\det{}_{TM}K(t)
\det{}_{TM}\left[1+t^2C(t)P\right]
\nonumber\\[5pt]
&&\times
\det{}_{TM}\left\{1+t^2[E(t)-S(t)K^{-1}(t)S(t)]P\right\}
\,,
\\[10pt]
\Psi(t)&=&\left(\Psi^\m_{\ \n}(t)\right)
=\left[1+t^2C(t)P\right]^{-1}C(t)\,.
\eea

The formula (\ref{1500}) exhibits the general structure of the heat
trace. One sees immediately how the endomorphism $Q$ and its
first derivatives $\nabla Q$ enter the result. 
The nontrivial information is
contained only in a
scalar, $\Phi(t)$, and a tensor, $\Psi_{\m\n}(t)$.
These objects are
constructed purely from the curvature $\R_{\m\n}$ and the
second derivatives of the endomorphism $Q$, $\nabla\nabla Q$.
Thus, the heat kernel coefficients $A_k$ are
constructed from three different types of scalar (connected) blocks,
$Q$, $\Phi_{(n)}(\R, \nabla\nabla Q)$ and
$\nabla_\m Q\Psi^{\m\n}_{(n)}(\R, \nabla\nabla Q)\nabla_\n Q$.
They are listed explicitly up to $A_8$ in \cite{avramidi95c}.


\subsection{Homogeneous Bundles over Symmetric Spaces}

The exposition in this section closely follows our papers
\cite{avramidi94a,avramidi96,avramidi08a,avramidi08b}.
Our goal is to compute the heat kernel of the Laplace type operator 
$L=-\Delta+Q$ in the zero-order of the low-energy approximation.
The difference with the previous sections is that now we are going to do it on most general covariantly constant background, 
that is, bundles with parallel curvature (that are called 
{\it homogeneous bundles}) on Riemannian manifolds with parallel curvature
(that are
called {\it symmetric spaces}).

It is well known that heat invariants are  determined essentially by
local geometry. They are polynomial invariants in the curvature with
universal constants that do not depend on the global properties of the
manifold \cite{gilkey95}. It is this universal structure that we are
interested in this paper.   Our goal is to compute the heat kernel
asymptotics of the Laplacian acting on homogeneous vector bundles over
symmetric spaces.

In this section we will further  assume that $M$ is a {\it
locally
symmetric space} with a Riemannian metric with the parallel curvature
\be
\nabla_\mu R_{\alpha\beta\gamma\delta}=0\,,
\ee
which means, in particular, that the curvature satisfies the 
integrability constraints
\be
R^{fg}{}_{ e a}R^e{}_{b cd} 
-R^{fg}{}_{ e b}R^e{}_{a cd} 
+ R^{fg}{}_{ ec}R^e{}_{d ab}
-R^{fg}{}_{ ed}R^e{}_{c ab}
= 0\,.
\label{312}
\ee
In the following we will also consider {\it homogeneous vector bundles} with
parallel bundle curvature
\be
\nabla_\mu\mathcal{R}_{\alpha\beta}=0\,,
\ee
which means that the curvature satisfies the integrability constraints
\be 
[\mathcal{R}_{cd},\mathcal{R}_{ab}]
-R^f{}_{acd}\mathcal{R}_{fb} -
R^f{}_{bcd}\mathcal{R}_{af} = 0\,. 
\label{227} 
\ee
Finally, we consider a parallel section $Q$ of the endomorphism bundle
$\End({\cal V})$, that is,
\be
\nabla_\mu Q=0\,,
\ee
which means that
\be 
[\mathcal{R}_{cd},Q] = 0\,. 
\label{227xx} 
\ee


We will use normal coordinates defined above.
Note that for symmetric
spaces normal coordinates cover the whole manifold except for a set of
measure zero where they become singular \cite{camporesi90}. This set is
precisely the set of points conjugate to the fixed point $x'$ (where 
$\Delta^{-1}(x,x')=0$) and of points that can be connected to the point
$x'$ by multiple geodesics. In any case, this set is a set of measure
zero and, as we will show below, it can be dealt with  by some
regularization technique. Thus, we will use the normal coordinates
defined above for the whole manifold. 

\subsubsection{Curvature Group of a Symmetric Space}

We assumed that the manifold $M$ is locally symmetric. Since we also
assume that it is simply
connected and complete, it is a globally symmetric space (or simply
symmetric space).  A symmetric space is said to be
compact, non-compact or Euclidean if all sectional curvatures are
positive, negative or zero. A generic symmetric space has the structure
$
M=M_0\times M_s\,,
$
where $M_0=\RR^{n_0}$ and
$M_s$ is a semi-simple symmetric space; it is a product
of a compact symmetric space $M_+$ and a non-sompact symmetric space $M_-$,
$
M_s=M_+\times M_-\,.
$
Of course, the dimensions must satisfy the relation  $n_0+n_s=n$,
where $n_s=\dim M_s$.

Let $\Lambda_2$ be the vector space of $2$-forms on $M$
at a fixed point $x'$. It has the
dimension $\dim \Lambda_2=n(n-1)/2$, and the inner product in
$\Lambda_2$ is defined by 
\be
\left<X,Y\right>=\frac{1}{2}X_{ab}Y^{ab}\,.
\ee
The Riemann curvature tensor naturally defines the curvature operator 
\be
{\rm Riem}: \Lambda_2\to \Lambda_2
\ee
by
\be
({\rm Riem}\, X)_{ab}=\frac{1}{2}R_{ab}{}^{cd}X_{cd}\,.
\ee
This operator is symmetric and has real eigenvalues which determine the
principal sectional curvatures. Now, let ${\rm Ker}\,({\rm Riem})$
and ${\rm Im}\,({\rm Riem})$ be the
kernel and the range 
of this operator and 
\be
p=\dim {\rm Im}({\rm Riem})
=\frac{n(n-1)}{2}-\dim {\rm Ker}\,({\rm Riem})\,\,.
\ee
Further, let $\lambda_i$, $(i=1,\dots, p)$, be the non-zero eigenvalues,
and $E^i{}_{ab}$ be the corresponding orthonormal eigen-two-forms.
Then the components of the curvature tensor
can be presented in the form \cite{avramidi96}
\be
R_{abcd} = \beta_{ik}E^i{}_{ab}E^k{}_{cd}\,,
\label{236}
\ee
where  $\beta_{ik}$ is a symmetric, in fact, diagonal, nondegenerate  
$p\times p$ matrix.
Of course, the zero eigenvalues of the curvature operator correspond to
the flat subspace $M_0$, the positive ones correspond to the compact
submanifold $M_+$ and the negative ones to the non-compact submanifold
$M_-$. Therefore, ${\rm Im}\,({\rm Riem})=T_x M_s$.

In the following the
Latin indices from the middle of the alphabet will be used to denote
tensors in ${\rm Im}({\rm Riem})$; 
they should not be confused with the Latin indices from
the beginning of the alphabet which denote tensors in $M$.
They will be raised and lowered with
the matrix $\beta_{ik}$ and its inverse
$\beta^{ik}$.

Next, we define the traceless $n\times n$
matrices $D_i=(D^a{}_{ib})$,
where
\be
D^a{}_{ib}=-\beta_{ik}E^k{}_{cb}\delta^{ca}\,.
\ee
The matrices $D_i$ are known to be the generators of the  holonomy
algebra, $\mathcal{H}$, i.e. the Lie algebra of the restricted holonomy 
group, $H$,
\be
[D_i, D_k] = F^j{}_{ik} D_j\,,
\label{310}
\ee
where $F^j{}_{ik}$ are the structure constants of the holonomy group.
The structure constants of the holonomy group define the
$p\times p$ matrices $F_i$, by $(F_i)^j{}_k=F^j{}_{ik}$, which generate
the adjoint representation  of the holonomy algebra,
\be
[F_i, F_k] = F^j{}_{ik} F_j\,.
\ee

For symmetric spaces the introduced quantities satisfy additional
algebraic constraints. The most important consequence of the eq.
(\ref{312}) is the equation
\cite{avramidi96}
\be
E^i{}_{a c} D^c{}_{kb}
-E^i{}_{b c} D^c{}_{ka} 
= F^i{}_{kj}E^j{}_{ab}\,.
\label{313}
\ee

Now, we introduce a new type of indices, the capital Latin indices,
$A,B,C,\dots,$ which split according to $A=(a,i)$ and run from $1$ to
$N=p+n$. We define new quantities $C^A{}_{BC}$ by
\be
C^i{}_{ab}=E^i{}_{ab}, \qquad 
C^a{}_{ib}=-C^a{}_{bi}=D^a{}_{ib}, \qquad 
C^i{}_{kl}=F^i{}_{kl}\,,
\label{317}
\ee
all other components being zero.
Let us also introduce rectangular $p\times n$ matrices 
$T_a$ by $(T_a)^j{}_c=E^j{}_{ac}$ and the $n\times p$ matrices
$\bar T_a$ by $(\bar T_a)^b{}_i=-D^b{}_{ia}$. 
Then we can define $N\times N$ matrices $C_A=(C_a,C_i)$
\be
C_a = \left(
\begin{array}{cc}
0 & \bar T_a \\
T_a & 0 \\
\end{array}
\right)\,,
\qquad
C_i = \left(
\begin{array}{cc}
D_i & 0 \\             
0 & F_i\\
\end{array}
\right),
\label{318}
\ee
so that $(C_A)^B{}_C=C^B{}_{AC}$. 

Then one can prove the following \cite{avramidi96}:
\begin{theorem}
The matrices $C_A$ generate the adjoint representation of a Lie algebra
$\mathcal{G}$ 
with the structure constants $C^A{}_{BC}$, that is,
\be
[C_A, C_B]=C^C{}_{AB}C_C\,,
\label{320}     
\ee
\end{theorem}

For the lack of a better name we call the algebra $\mathcal{G}$ the {\it
curvature algebra}. As it will be clear from the next section it is a
subalgebra of the total isometry algebra of the symmetric space. It
should be clear  that the holonomy algebra $\mathcal{H}$ is the subalgebra
of the curvature algebra  $\mathcal{G}$.  
The curvature algebra $\mathcal{G}$ is compact; it is a direct sum
of two ideals, 
$
\mathcal{G}=\mathcal{G}_0\oplus \mathcal{G}_s,
\label{290xx}
$
an Abelian center $\mathcal{G}_0$ of dimension $n_0$ 
and a semi-simple algebra $\mathcal{G}_s$ of
dimension $p+n_s$.

Next, we define a symmetric nondegenerate
$N\times N$ matrix
\be
(\gamma_{AB}) = 
\left(
\begin{array}{cc}
\delta_{ab} & 0 \\
0 & \beta_{ik} \\
\end{array}
\right)
\,.
\label{319}
\ee
This matrix and its inverse 
$\gamma^{AB}$
will be
used to lower and to raise the capital Latin indices. 

\subsubsection{Killing Vectors Fields}

We will use extensively the isometries of the symmetric space $M$. 
We follow the approach developed in 
\cite{avramidi96,avramidi91,avramidi00,avramidi08b}.
The
generators of isometries are the Killing vector fields
$\xi$.
The set of all Killing vector fields forms a representation of the 
isometry algebra, the Lie algebra of the isometry group
of the manifold $M$. We define two
subspaces of the isometry algebra. One subspace is formed by Killing
vectors (called translations) satisfying the initial
conditions 
$
\nabla_\mu\xi^\nu\big|_{x=x'}=0\,,
\label{289}
$
and another subspace is formed by Killing vectors
(called rotations) 
satisfying the
initial conditions
$
\xi^\nu\big|_{x=x'}=0\,.
\label{290}
$

One can easily show that a basis of translations 
can be chosen as
\be
P_a=\left(\sqrt{K}\cot\sqrt{K}\right)^b{}_a\frac{\partial}{\partial y^b}\,.
\label{226}
\ee
where $K=(K^a{}_b)$ is a matrix defined by
\be
K^a{}_b=R^a{}_{cbd}y^cy^d\,.
\ee
We can also show that the vector fields
\be
L_i=-D^b{}_{ia}y^a\frac{\partial}{\partial y^b}\,
\label{229}
\ee
define $p$ linearly independent rotations.
By adding the trivial Killing vectors for flat subspaces we find that
the number of independent rotations is
$
p+n_0n_s+n_0(n_0-1)/2\,.
$
We introduce the following notation $(\xi_A)=(P_a,L_i)$. 

By using the explicit
form of the Killing vector fields obtained above 
\cite{avramidi96} one can prove the following theorem.
\begin{theorem}
The Killing vector fields $\xi_A$ form 
form a representation of the
curvature algebra $\mathcal{G}$
\be
[\xi_A, \xi_B]=C^C{}_{AB}\xi_C\,.
\label{320a}
\ee
\end{theorem}
Notice that they {\it do not} generate  the complete isometry algebra of
the symmetric space $M$. The 
curvature algebra $\mathcal{G}$ introduced in the previous section
is a subalgebra of the total isometry algebra.
It is clear that the Killing vector fields $L_i$ form a
representation of the holonomy algebra $\mathcal{H}$, which is the isotropy
algebra of the semi-simple submanifold $M_s$, and a  subalgebra of the
total isotropy algebra of the symmetric space $M$.

\subsubsection{Homogeneous Vector Bundles}

Let $h^a{}_b$ be the projection 
to
the subspace $T_x M_s$ of the tangent space
and
\be
q^a{}_{b}=\delta^a{}_b-h^a{}_b\,
\ee
be the projection tensor to the flat subspace $\RR^{n_0}$.
Since the curvature exists only in the semi-simple submanifold $M_s$, the
components of the curvature tensor $R_{abcd}$, as well as the tensors
$E^i{}_{ab}$,  are non-zero only in the semi-simple subspace $T_xM_s$.
Then
\be
R_{abcd}q^a{}_e=R_{ab}q^a{}_e=E^i{}_{ab}q^a{}_e=D^a{}_{ib}q^b{}_e
=D^a{}_{ib} q_a{}^e=0\,.
\ee 

Equation (\ref{227}) imposes strong constraints on the curvature of the
homogeneous bundle $\mathcal{W}$. We define
\be
\mathcal{B}_{ab}=\mathcal{R}^{YM}_{cd}q^c{}_a q^d{}_b\,,
\qquad
\mathcal{E}_{ab}=\mathcal{R}^{YM}_{cd}h^c{}_bh^d{}_b\,,
\label{2139}
\ee
so that
\be
{\cal R}^{YM}_{ab}=\mathcal{E}_{ab}+\mathcal{B}_{ab}\,.
\ee
Then, from eq. (\ref{227}) we obtain
\be
[\mathcal{B}_{ab}, \mathcal{B}_{cd}]
=[\mathcal{B}_{ab}, \mathcal{E}_{cd}]=0\,,
\ee
and
\be
[\mathcal{E}_{cd}, \mathcal{E}_{ab}] 
- R^f{}_{acd}\mathcal{E}_{fb}
- R^f{}_{bcd}\mathcal{E}_{af}
= 0\,.
\label{227a}
\ee
This means that  $\mathcal{B}_{ab}$ takes values in an Abelian ideal of the
gauge algebra $\mathcal{G}_{YM}$ and $\mathcal{E}_{ab}$ takes values in the
holonomy algebra. More precisely, eq. (\ref{227a}) is only possible if
the holonomy algebra $\mathcal{H}$ is an ideal of the gauge algebra 
$\mathcal{G}_{YM}$. Thus, the gauge group $G_{YM}$ must have a subgroup
$Z\times H$, where $Z$ is an Abelian group and $H$ is the holonomy
group.

Let $X_{ab}$ be the
generators of the orthogonal algebra $\mathcal{SO}(n)$ is some
representation $X$. Then the matrices 
$T_i=-\frac{1}{2}D^a{}_{ib}X^b{}_a$ are the generators of the gauge
algebra ${\cal G}_{YM}$ realizing a representation $T$
of the 
holonomy algebra ${\cal H}$.
Next, we can show that
the
curvature of the homogeneous bundle $\mathcal{W}$
is given by
\bea
\mathcal{R}^{YM}_{ab}&=&
-E^i{}_{ab}T_i+\mathcal{B}_{ab}
=\frac{1}{2}R^{cd}{}_{ab}X_{cd}+\mathcal{B}_{ab}
\,.
\label{2153mm}
\eea

Now, we consider the representation $\Sigma$ of the orthogonal algebra
${\cal SO}(n)$
defining the spin-tensor bundle $\mathcal{T}$ and define the matrices
\be
G_{ab}=\Sigma_{ab}\otimes\II_X+\II_\Sigma\otimes X_{ab}\,.
\ee
Obviously, these matrices are the generators of the orthogonal algebra
${\cal SO}(n)$
in the product representation $\Sigma\otimes X$.
Next, the matrices $Y_i=-\frac{1}{2}D^a{}_{ib}\Sigma^b{}_a$
form a representation $Y$ of the holonomy algebra ${\cal H}$
and the matrices
\be
{\cal R}_i=-\frac{1}{2}D^a{}_{ib}G^b{}_a
\ee
are the generators of the holonomy algebra in the product
representation ${\cal R}=Y\otimes T$.

Then the total curvature,
that is, the commutator of covariant derivatives, (\ref{220}) of  a
twisted spin-tensor bundle $\mathcal{V}$ is
\bea
{\cal R}_{ab}&=&
-E^i{}_{ab}{\cal R}_i+{\cal B}_{ab}
=
\frac{1}{2}R^{cd}{}_{ab}G_{cd}+\mathcal{B}_{ab}
\,.
\label{2144}
\eea

\subsubsection{Twisted Lie Derivatives}

Let $\varphi$ be a section of a  twisted homogeneous spin-tensor bundle
$\mathcal{T}$. Let $\xi_A$ be the basis of Killing vector fields. Then the
covariant (or generalized, or twisted)  
Lie derivative of $\varphi$ along $\xi_A$ is defined by
\be
\mathcal{L}_A\varphi=
\left(\xi_A{}^\mu\nabla_\mu
+\frac{1}{2}\xi_A{}^a{}_{;b}G^b{}_a
\right)\varphi\,.
\label{2160x}
\ee

One can prove the theorem \cite{avramidi08a,avramidi08b}.
\begin{theorem}
The operators $\mathcal{L}_A$ 
satisfy the commutation relations
\bea
[\mathcal{L}_A, \mathcal{L}_B]
&=&C^C{}_{AB} \mathcal{L}_C+\mathcal{B}_{AB},
\label{2173xx}
\eea
where
\be
\mathcal{B}_{AB}=\left(
\begin{array}{cc}
\mathcal{B}_{ab} & 0\\
0 & 0\\
\end{array}
\right)\,,
\label{2171xx}
\ee
\end{theorem}

The operators $\mathcal{L}_A$ form an algebra that
is a direct sum of a nilpotent ideal and a 
semisimple algebra. For the lack of a better name we call this algebra
{\it gauged curvature algebra} and denote it by $\mathcal{G}_{\rm gauge}$.

Now, let us define the operator
\be
{\cal L}^2=\gamma^{AB}{\cal L}_A{\cal L}_B
\ee
and the Casimir operator of the holonomy group
\be
{\cal R}^2=\frac{1}{4}R^{abcd}G_{ab}G_{cd}\,.
\ee

Then one can prove that \cite{avramidi08b}
\begin{theorem}
The Laplacian $\Delta$ acting on sections of a twisted spin-tensor
bundle $\mathcal{V}$ over a symmetric space has the form
\be
\Delta=\mathcal{L}^2-\mathcal{R}^2\,.
\label{2110}
\ee
\end{theorem}

\subsubsection{Geometry of the Curvature Group}

Let $G_{\rm gauge}$ be the gauged curvature group and $H$ be its
holonomy subgroup.  Both these groups have compact algebras. However,
while the holonomy group is always compact, the curvature group is, in
general, a product of a nilpotent group, $G_0$, and a
semi-simple group, $G_s$,
$
G_{\rm gauge}=G_0\times G_s\,.
$
The semi-simple group $G_s$ is a product $G_s=G_+\times G_-$ of a
compact $G_+$ and a non-compact $G_-$ subgroups.

Let $\xi_A$ be the basis Killing vectors, $k^A$ be the canonical
coordinates on the curvature group $G$ and $\xi(k)=k^A\xi_A$.
The canonical coordinates are exactly the normal coordinates
on the group defined above.
Let $C_A$ be the generators of the curvature group in
adjoint representation and $C(k)=k^AC_A$. 

Let $X=(X_A{}^M)$ be the matrix defined by
\be
X=\frac{C(k)}{1-\exp[-C(k)]}\,,
\ee
and $X_A$ be the vector fields
on the group $G$ defined by
\be
X_A=X_A{}^M\frac{\partial}{\partial k^M}\,.
\ee
Then one can show that \cite{avramidi08b}
the vector fields $X_A$ form a 
representation of the curvature algebra ${\cal G}$
\be
[X_A,X_B]=C^C{}_{AB}X_C\,.
\label{37xx}
\ee
The vector fields $X_A$ are nothing but the right-invariant
vector fields.

Since we will actually be working with the gauged curvature group,
we introduce now the operators ({\it covariant right-invariant
vector fields}) $J_A$ by
\be
J_A=X_A-\frac{1}{2}\mathcal{B}_{AB}k^B\,,
\ee
Then we show \cite{avramidi08b}
that the operators $J_A$ form the following algebra
\be
[J_A,J_B]=C^C{}_{AB}J_C+\mathcal{B}_{AB}\,.
\label{326xx}
\ee
Thus, the operators $J_A$ form a representation of the gauged curvature
algebra ${\cal G}_{\rm gauge}$.

Now, let $\mathcal{L}_A$ be the Lie derivatives
and
$\mathcal{L}(k)=k^A \mathcal{L}_A$. 
Then we find \cite{avramidi08b}
\be
J_A \exp[\mathcal{L}(k)]=\exp[\mathcal{L}(k)]\mathcal{L}_A\,.
\label{jtolxxx}
\ee
Notice that $J_A$ are first order differential operators with respect to
$k^A$, whereas ${\cal L}_A$ are first-order partial differential operators
with respect to the coordinates $x$ acting on sections of the bundle 
${\cal V}$.

\subsubsection{Heat Kernel on the Curvature Group}

Now, let us define the operator
\be
J^2=\gamma^{AB}J_AJ_B
\ee
and the invariant (scalar curvature of the curvature group) 
\be
R_G=-\frac{1}{4}\gamma^{AB}C^C{}_{AD}C^D{}_{BC}\,.
\label{312xx}
\ee

Then by using the properties of the right-invariant vector fields
$J_A$ one can find the heat kernel of the operator $J^2$ on the 
curvature group ${\cal G}$ \cite{avramidi08b}.
\begin{theorem}
Let $\Phi(t;k)$ be a function on the curvature 
group defined 
in canonical coordinates $k^A$ by
\bea
\Phi(t;k)&=&(4\pi t)^{-N/2}
\left[\det{}_{TM}
\left(\frac{\sinh\left[t\mathcal{B}\right]}
{t\mathcal{B}}\right)\right]^{-1/2}
\left[\det{}_\mathcal{G}
\left(\frac{\sinh\left[C(k)/2\right]}
{C(k)/2}\right)\right]^{-1/2}
\nonumber\\
&&
\times
\exp\left(
-\frac{1}{4t}\left<k,\gamma t\mathcal{B}
\coth(t\mathcal{B})k\right>
+\frac{1}{6}R_G t\right)\,,
\label{46a}
\eea
where
$\left<u,\gamma v\right>=\gamma_{AB}u^Av^B$ is the inner product on
the algebra $\mathcal{G}$.
Then $\Phi(t;k)$ satisfies the heat equation
\be
\partial_t \Phi = J^2\Phi\,,
\label{423}
\ee
and the initial condition
\be
\Phi(0;k)=\gamma^{-1/2}\delta(k)\,,
\label{428}
\ee
where $\gamma=\det \gamma_{AB}$.
\end{theorem}


In the following we will complexify the gauged  curvature group in the
following sense. We extend the  canonical coordinates
$(k^A)=(p^a,\omega^i)$ to the whole complex Euclidean space  $\CC^N$.
Then all group-theoretic functions introduced above become analytic
functions of $k^A$ possibly with some poles on the real section $\RR^N$
for compact groups.  In fact, we replace the actual real slice $\RR^N$
of $\CC^N$ with an $N$-dimensional subspace $\RR^N_{\rm reg}$ in $\CC^N$
obtained by rotating the real section $\RR^N$ counterclockwise in
$\CC^N$ by $\pi/4$. That is, we replace each coordinate $k^A$ by
$e^{i\pi/4}k^A$. In the complex domain the group becomes non-compact. We
call this procedure the {\it decompactification}. If the group is
compact, or has a compact subgroup, then this plane will cover the
original group infinitely many times. 

Since the metric $(\gamma_{AB})=\diag(\delta_{ab},\beta_{ij})$ is not
necessarily positive definite,  (actually, only the metric of the
holonomy group $\beta_{ij}$ is non-definite) we analytically continue
the function $\Phi(t;k)$ in the  complex plane of $t$ with a cut along
the negative imaginary axis so that
$-\pi/2<\arg\,t<3\pi/2$.  Thus, the function $\Phi(t;k)$ defines an
analytic function of $t$ and $k^A$. For the purpose of the following
exposition we shall consider $t$ to be {\it real negative}, $t<0$. This
is needed in order to make all integrals convergent and well defined and
to be able to do the analytical continuation.

As we will show below, the singularities occur only in the holonomy
group. This means that there is no need to complexify the coordinates
$p^a$. Thus, in the following we assume the coordinates $p^a$ to be real
and the coordinates $\omega^i$ to be complex, more precisely, to take
values in the $p$-dimensional subspace $\RR^p_{\rm reg}$ of $\CC^p$ 
obtained by rotating $\RR^p$ counterclockwise by $\pi/4$ in $\CC^p$ That
is, we have $\RR^N_{\rm reg}=\RR^n\times \RR^p_{\rm reg}$.

This procedure (that we call a regularization) with the nonstandard
contour of integration is necessary  for the convergence of the
integrals below since we are treating both the compact and the
non-compact symmetric spaces simultaneously. Remember, that, in general,
the nondegenerate diagonal matrix $\beta_{ij}$ is not positive definite.
The space $\RR^p_{\rm reg}$ is chosen in such a way to make the Gaussian
exponent purely imaginary. Then the indefiniteness of the matrix $\beta$
does not cause any problems. Moreover, the integrand does not have any
singularities on these contours. The convergence of the integral is
guaranteed by the exponential growth of the sine for imaginary argument.
These integrals can be computed then in the following way. The
coordinates $\omega^j$ corresponding to the compact directions  are
rotated further by another $\pi/4$ to imaginary axis and the coordinates
$\omega^j$ corresponding to the non-compact directions are rotated back
to the real axis. Then, for $t<0$ all the integrals below are well
defined and convergent and define an analytic function of $t$ in a complex
plane with a cut along the negative imaginary axis.

\subsubsection{Heat Trace}

Now, by using the heat kernel (\ref{46a})
of the operator $J^2$ on the curvature group obtained above,
the relation (\ref{2110}) of the Laplacian and the operator ${\cal L}^2$,
and the property (\ref{jtolxxx}) one can find the following integral
representation of the heat semigroup of the Laplace-type operator
\cite{avramidi08b}.
\begin{theorem}
Let $L=-\Delta+Q$ be the Laplace type operator
acting on sections of a homogeneous
twisted spin-tensor vector bundle over a symmetric space. Then the
heat semigroup $\exp(-tL)$
can be represented in form of an integral
\bea
\exp(-tL) 
&=& 
(4\pi t)^{-N/2}
\left[\det{}_{TM}\left(\frac{
\sinh(t\mathcal{B})}{t\mathcal{B}}\right)\right]^{-1/2} 
\exp\left(-tQ-t\mathcal{R}^2+ {1\over 6} R_G t\right)
\nonumber\\
&&
\times\int\limits_{\RR^N_{\rm reg}} dk\; \gamma^{1/2}
\left[\det{}_\mathcal{G}
\left({\sinh[C(k)/2]\over C(k)/2}\right)\right]^{1/2}
\nonumber\\
&&
\times
\exp\left\{ -{1\over 4t}
\left<k,\gamma t\mathcal{B}\coth(t\mathcal{B})k\right>
\right\}
\exp[\mathcal{L}(k)]\,.
\label{49}
\eea
\end{theorem}


The heat trace can be obtained by acting by the heat
semigroup $\exp(-tL)$ on the delta-function,
To be able to use this integral representation we need to compute the
action of the isometries $\exp[\mathcal{L}(k)]$ on the delta-function.


Let $\omega^i$ be the canonical coordinates on the holonomy
group $H$ and $(k^A)=(p^a,\omega^i)$ be the natural splitting of the
canonical coordinates on the curvature group $G$.
Then we can prove that \cite{avramidi08b}
\be
\left[\exp[\mathcal{L}(k)]\delta(x,x')\right]^{\rm diag}
=\left[\det{}_{TM}\left(\frac{\sinh[D(\omega)/2]}{D(\omega)/2}
\right)\right]^{-1}
\exp[\mathcal{R}(\omega)]
\delta(p)\,,
\label{424a}
\ee
where 
$D(\omega)=\omega^iD_i$ and
${\cal R}(\omega)=\omega^i{\cal R}_i$.

We implicitly assumed that there are no closed geodesics
and that the equation of closed orbits of isometries
has a unique solution. On compact symmetric
spaces this is not true: there are infinitely many closed geodesics and
infinitely many closed orbits of isometries.   However, these global
solutions, which reflect the global topological structure of the
manifold, will not affect our local analysis.  In particular, they do
not affect the asymptotics of the heat kernel.  That is why, we have
neglected them here.  This is reflected in the fact that the Jacobian in
(\ref{424a}) can become singular when the coordinates of the holonomy
group $\omega^i$ vary from $-\infty$ to $\infty$.  Note that the exact
results for compact symmetric spaces can be obtained by an analytic
continuation from the dual noncompact case when such closed geodesics
are absent \cite{camporesi90}. That is why we proposed above to
complexify our holonomy group. If the coordinates $\omega^i$ are complex
taking values in the subspace $\RR^p_{\rm reg}$ defined above, then the
equation of closed orbits should have a unique 
trivial solution and the Jacobian is
an analytic function. It is worth stressing once again  that the
canonical coordinates cover the whole group except for a set of measure
zero.  Also a compact subgroup is covered  infinitely many times. 

Now by using the above lemmas and the theorem we can compute the heat
trace. 
We define the invariant (scalar curvature of the holonomy group)
\be
R_H=-\frac{1}{4}\beta^{ij}F^k{}_{il}F^l{}_{jk}\,.
\ee

\begin{theorem}
The heat trace of the operator $L$
has the form
\bea
\Theta(t)
&=&(4\pi t)^{-n/2}
\int\limits_M d\vol\;\tr_V
\left[\det{}_{TM}\left(\frac{
\sinh(t\mathcal{B})}{t\mathcal{B}}\right)\right]^{-1/2}
\exp\left\{\left({1\over 8} R + {1\over 6} R_H 
-\mathcal{R}^2-Q\right)t\right\}
\nonumber
\label{437a}
\\
&&
\times
\int\limits_{\RR^n_{\rm reg}} 
\frac{d\omega}{(4\pi t)^{p/2}}\;\beta^{1/2}
\exp\left\{-{1\over 4 t}\left<\omega,\beta\omega\right>\right\}
\cosh\left[\,\mathcal{R}(\omega)\right]
\nonumber\\
&& 
\times\left[\det{}_\mathcal{H}
\left({\sinh\left[\,F(\omega)/2\right]\over 
\,F(\omega)/2}\right)\right]^{1/2}
\left[\det{}_{TM}\left({\sinh\left[\,D(\omega)/2\right]\over 
\,D(\omega)/2}\right)\right]^{-1/2}\,,
\eea
where $\beta=\det \beta_{ij}$,
$\left<\omega,\beta\omega\right>=\beta_{ij}\omega^i\omega^j$
and $F(\omega)=\omega^i F_i$.

\end{theorem}

This equation can be used now to generate all heat kernel
coefficients $A_k$ for any locally symmetric space  simply by expanding
it in a power  series in $t$. By using the standard Gaussian averages
one can obtain now all heat kernel coefficients in terms of 
traces of various contractions 
of the matrices $D^a{}_{ib}$ and $F^j{}_{ik}$ with  the
matrix  $\beta^{ik}$.
All these quantities are curvature invariants and
can be  expressed directly in terms of the Riemann tensor.

\section{Low Energy Effective Action in Quantum General Relativity}
\setcounter{equation}0

We can apply now the obtained results for the heat trace to compute
the low-energy one-loop effective action in quantum general relativity
given by (\ref{eaxxz}).
In the Euclidean formulation we have
\be
\Gamma_{(1)}=\frac{1}{2}\left(\log\Det \hat L-2\log\Det F\right)\,,
\ee
which, in the zeta regularization takes the form
\be
\Gamma_{(1)}=-\frac{1}{2}
\left(\zeta'_{\hat L}(0)-2\zeta'_F(0)\right)\,,
\ee
where $\zeta_{\hat L}(s)$ and $\zeta_F(s)$ are the zeta
functions of the graviton operator $\hat L$ and the ghost
operator $F$.
Now, let us define the total zeta function by
\be
\zeta_{GR}(s)=\zeta_{\hat L}(s)-2\zeta_F(s)\,.
\ee
Then the effective action is
\be
\Gamma_{(1)}=-\frac{1}{2}\zeta'_{GR}(0)\,.
\ee
Next, by using the definition of the zeta function we obtain
\be
\zeta_{GR}(s)=\frac{\mu^{2s}}{\Gamma(s)}
\int\limits_0^\infty dt\; t^{s-1}e^{t\lambda}
\Theta_{GR}(t)\,,
\ee
where 
\be
\Theta_{GR}(t)=
\Theta_{\hat L}(t)
-2\Theta_F(t)\,,
\ee
$\mu$ is a renormalization parameter introduced to preserve
dimensions and $\Theta_{\hat L}(t)$ and $\Theta_F(t)$ are the heat
traces of the operators $\hat L$ and $F$.
Here $\lambda$ is a sufficiently large negative infrared cutoff parameter
introduced
to regularize any infrared divergences which are present if 
the operators $\hat L$ and $F$ have negative modes. The parameter
$\lambda$ should be
set to zero at the end of the calculations.

Now, notice that both operators $\hat L$ and $F$ are of Laplace type,
that is, $-\Delta+Q$, acting on pure tensor bundles;
so, there is
no Yang-Mills group here, 
$\tilde{\cal R}_{ab}={\cal E}_{ab}={\cal B}_{ab}=0$. 
The operator $\hat L$ acts on the bundle ${\cal T}_{(2)}
=T^*M\otimes T^*M$
of symmetric two-tensors 
and the operator $F$ acts on sections of the tangent
bundle ${\cal T}_{(1)}=TM$.
The potentials,
$Q$,
for both operators are obviously read off from their definition
\be
\left(Q_{(1)}\right)^a{}_b=-R^a{}_b\,,
\ee
\bea
\left(Q_{(2)}\right)_{cd}{}^{ab}
&=&
-2R^{(a}{}_c{}^{b)}{}_d
-2\delta^{(a}{}_{(c}R^{b)}{}_{d)}
+R_{cd}g^{ab}
+\frac{2}{n-2}g_{cd}R^{ab}
\nonumber\\
&&
-\frac{1}{(n-2)}g_{cd}g^{ab}R
+\delta^a{}_{(c}\delta^b{}_{d)}(R-2\Lambda)
\,.
\eea

The generators of the orthogonal group $SO(n)$ in the
vector representation are
\be
\left(\Sigma_{(1)}{}_{ab}\right)^c{}_d=2\delta^{c}{}_{[a}g_{b]d}\,.
\ee
The generators of the orthogonal group $SO(n)$ in the
symmetric $2$-tensor representation are
\be
\left(\Sigma_{(2)}{}_{ab}\right){}_{cd}{}^{ef}
=-4\delta^{(e}{}_{[a}g_{b](d}\delta^{f)}{}_{c)}
\,.
\ee
The generators of the holonomy group are
\be
{\cal R}_{(1)}{}_i=D_{i}\,,
\ee
and
\be
{\cal R}_{(2)}{}_i
=-2 D_{i}\vee I_{(1)}\,,
\ee
which, in component language, reads
\be
\left({\cal R}_{(1)}{}_i\right)^a{}_b=
D^a{}_{ib}\,,
\ee
and
\be
\left({\cal R}_{(2)}{}_i\right)_{cd}{}^{ab}
=-2 D^{(a}{}_{i(d}\delta^{b)}{}_{c)}\,.
\ee
The Casimir operators are
\be
\left({\cal R}^2_{(1)}\right)^a{}_b
=-R^a{}_{b}\,,
\ee
and
\be
\left({\cal R}^2_{(2)}\right)_{cd}{}^{ab}
=2R^{(a}{}_{d}{}^{b)}{}_{c}
-2\delta^{(a}{}_{(c}R^{b)}{}_{d)}
\,.
\ee

By using the results for the heat traces described above we
obtain the total heat trace
\bea
\Theta_{GR}(t)
&=&(4\pi t)^{-n/2}
\int\limits_M d\vol\;
\exp\left\{\left({1\over 8} R + {1\over 6} R_H\right)t\right\}
\label{437axxz}
\\
&&
\times
\int\limits_{\RR^n_{\rm reg}} 
\frac{d\omega}{(4\pi t)^{p/2}}\;\beta^{1/2}
\exp\left\{-{1\over 4 t}\left<\omega,\beta\omega\right>\right\}
\Psi(t;\omega)
\nonumber\\
&& 
\times
\left[\det{}_\mathcal{H}
\left({\sinh\left[\,F(\omega)/2\right]\over 
\,F(\omega)/2}\right)\right]^{1/2}
\left[\det{}_{TM}\left({\sinh\left[\,D(\omega)/2\right]\over 
\,D(\omega)/2}\right)\right]^{-1/2}
\eea
where
\bea
\Psi(t;\omega)
&=&\exp\left[-t(R-2\Lambda)\right]\tr_{{\cal T}_{(2)}}
\exp\left(tV_{(2)}\right)
\cosh\left[2D(\omega)\vee I_{(1)}\right]
\nonumber\\[5pt]
&&
-2\tr_{TM}
\exp\left(t V_{(1)}\right)
\cosh\left[\,D(\omega)\right]
\,,
\label{psixxx}
\eea
and the matrices $V_{(1)}$ and $V_{(2)}$ are defined by
\be
\left(V_{(1)}\right)^a{}_b=2R^a{}_b\,,
\ee
\be
\left(V_{(2)}\right)_{cd}{}^{ab}
=4\delta^{(a}{}_{(c}R^{b)}{}_{d)}
-R_{cd}g^{ab}
-\frac{2}{n-2}g_{cd}R^{ab}
+\frac{1}{(n-2)}g_{cd}g^{ab}R
\,.
\ee

One can go further and compute the function $\Psi(t;\omega)$ by finding the
eigenvalues of the endomorphisms $V_{(1)}$ and $V_{(2)}$. However, we will not
do it here and leave the answer in the form (\ref{psixxx}). By using the
obtained heat trace one can compute now the zeta function and then the
effective action. We would like to stress two points here. First of all,
quantum general relativity is a non-renormalizable theory. Therefeore, even if
one gets a final result via the zeta-regularization one should not take it too
seriously. Secondly, our results for the heat kernel and, hence, for the
effective action are essentially non-perturbative. They contain an
infinite series of Feynmann diagrams and cannot be obtained in any
perturbation theory. One could try now to use this result for the analysis
of the ground state in quantum gravity. But this is a rather ambitious program
for the future.


\subsubsection*{Acknowledgement}

I would like to thank the organizers of the
summer school ``New Paths Towards Quantum Gravity''
and the workshop 
``Quantum Gravity: An Assessment'',
Bernhelm Boo\ss-Bavnbek,
Giampiero Esposito and Matthias Lesch,
for their kind invitation and for the financial support.


\end{document}